\documentclass[twocolumn]{article} 
\usepackage{fullpage}

\usepackage{balance} 

\usepackage{caption}
\usepackage{graphicx}
\usepackage{amsmath}
\usepackage{amsthm}
\usepackage{booktabs}

\usepackage{amsmath,amsthm,amsfonts}
\usepackage[ruled,vlined,linesnumbered]{algorithm2e}
\usepackage{caption}
\usepackage{subcaption}
\newcommand{\llElse}[1]{{\let\par\relax\lElse{#1}}}

\newtheorem{theorem}{Theorem}
\newtheorem{lemma}{Lemma}

\newtheorem{proposition}{Proposition}

\newcommand{\cK}{{\mathcal{K}}}

\newcommand{\cD}{{\mathcal{D}}}
\newcommand{\cS}{{\mathcal{S}}}
\newcommand{\cF}{{\mathcal{F}}}
\usepackage{url}

\usepackage{xcolor}

\usepackage{mathtools}

\DeclarePairedDelimiter\floor{\lfloor}{\rfloor}

\newif\ifnotes\notestrue

\title{Safe Delivery of Critical Services in Areas with Volatile Security Situation via a Stackelberg Game Approach}

\author{Tien Mai, Arunesh Sinha\\
Singapore Management University\\
\{atmai,aruneshs\}@smu.edu.sg}
\date{}

\newcommand{\BibTeX}{\rm B\kern-.05em{\sc i\kern-.025em b}\kern-.08em\TeX}

\usepackage{fancyhdr}
\usepackage{natbib}

\begin{document}
\maketitle

\begin{abstract}
  Vaccine delivery in under-resourced locations with security risks is not just challenging but also life threatening. The current COVID pandemic and the need to vaccinate have added even more urgency to this issue. 
   Motivated by this problem, we propose a general framework to set-up limited temporary (vaccination) centers that balance physical security and desired (vaccine) service coverage with limited resources. We set-up the problem as a Stackelberg game between the centers operator (defender) and an adversary, where the set of centers is not fixed a priori but is part of the decision output. This results in a mixed combinatorial and continuous optimization problem. As part of our scalable approximation of this problem, we provide a fundamental contribution by identifying general duality conditions of switching max and min when both discrete and continuous variables are involved. We perform detailed experiments to show that the solution proposed is scalable in practice.
\end{abstract}


\pagestyle{fancy}
\fancyhead{}




\section{Introduction}

Vaccine delivery has always been a challenge in under-resourced parts of the world~\cite{zaffran2013imperative}. The problem is further aggravated by the threat of violence against vaccine providers in areas where the security situation is volatile~\cite{AP}. A safer vaccine delivery plan for such places can save lives, both via vaccination of the general population and physical protection of the front line vaccine providers. Specifically, vaccination drives in underdeveloped areas are often performed by setting up a \emph{limited} number of temporary vaccination centers. Motivated by this issue, we propose and study a \emph{general} framework to set-up such temporary centers that balance physical security requirements of the centers and achieve desired vaccination coverage.


Our \emph{first contribution} is a flexible model that allows choice of a small subset of centers to operate, along with a consideration of how to allocate security resources to the operational centers. Further, the framework allows for fairness constraints that ensure fairness in center allocation for different geographical regions as well as fairness in security allocation to operating centers. The model is set-up as a Stackelberg game between the centers operator (defender) and a bounded rational adversary who follows the Quantal Response (QR) model. The model takes inspiration from Stackelberg security games~\cite{tambe2011security} in the manner in which security resources are allocated and the center selection aspect is inspired by product assortment and pricing problems~\cite{Wang2012capacitated}. While inspired from these models from disparate areas, we believe this is a \emph{first} model that brings together security issues and subset of centers (targets) selection within one framework.

Our \emph{second contribution} is a hybrid algorithm that combines the best of two different approaches. The first of these two approaches is a Mixed Interger Linear Program (MILP) with guaranteed approximation, and the second is a polynomial time heuristic. All algorithms work by a sequence of modifications to the original optimization problem, starting with a binary search on the objective value. The first algorithm exploits properties of the problem at hand to convert a bi-linear MIP formulation to a MILP with guaranteed approximation, but, MILPs are not scalable in practice.
Hence, we design a heuristic where we use Lagrangian duality and reach a  sub-problem which is a max-min-max tri-level optimization with discrete variables for the outer max problem and continuous variables for the two inner min-max problems. 

Then, in a \emph{fundamental technical contribution}, we identify general conditions for minimax equality when the variables involved are discrete and continuous. While the conditions do not hold in rare cases for our problem, we use this result as a heuristic to transform our sub-problem to a min-max-max problem and then we present a polytime approximation for the transformed problem. 
The polynomial time approach provides for immense scalability in practice, with solutions close to the MILP approach in almost all cases. Finally, we show that the MILP and heuristic can be made to work together in a scalable hybrid approach with approximation guarantees for the output solution.

We conduct thorough experiments to analyze various aspects of our three approaches. We show that our main hybrid algorithm is scalable in practice (solving for up to 5000 potential centers within 1.5 minutes) and also much better than competing baseline approaches. 
While we are inspired by vaccine centers operation, our model and ideas can be applied to critical services problems such as operation of temporary health camps, exam centers, etc. in underdeveloped and security risk-prone areas or even existing security games work where a subset of targets can be chosen to be made unavailable. 
\emph{All missing proofs are in the appendix}.

To summarize, our main contributions are:
\begin{enumerate}
    \item A novel and flexible security game model that brings together elements of choosing a subset of targets to operate and security considerations under one umbrella.
    \item A fundamental minimax equality result with discrete and continuous variables that is of independent interest in optimization research.
   \item Our algorithm that uses the above result as well as exploits other properties to solve the game approximately in polynomial time.
    \item A thorough experimental analysis showing the much superior scalability of our algorithm against baselines. 
\end{enumerate}


\subsection{Related Work}

Our work takes inspiration from two main distinct topics:


\textbf{Product Assortment and Pricing.} Our problem 
is most closely related to   
the problem that combines product assortment and price optimization problem in which a subset of products is chosen from an assortment of products and simultaneously the prices of products is chosen with the goal of maximizing profit in a market where the buyers choose a product following the  QR~\cite{Wang2012capacitated} model. The relation to our problem stems from the analogy of set of products to set of centers and of continuous prices to security allocation. However, prior work~\cite{Wang2012capacitated} solves an unconstrained optimization (no constraints on prices), whereas we deal with a constrained optimization. 

Other works in this area optimize just the product assortment with fixed prices~\cite{davis2014assortment,desir2020capacitated}. 
In other works (still fixed prices), different models of buyers have been considered such as rational best responding buyers~\cite{10.1007/978-3-030-04612-5_15} and other discrete choice models~\cite{davis2014assortment,gallego2015general}. More further away, there is work on an online version of the product assortment problem using a multi-armed bandits formulation~\cite{agrawal2016near,cheung2017thompson} but again with fixed prices.


\textbf{Stackelberg Security Games.} 
There is a large body of work in game theoretic models and algorithms for physical security~\cite{tambe2011security,xu2016mysteries,fang2017paws,sinha2018stackelberg,ijcai2020-35,yang2012computing,1501625} of a given fixed set of vulnerable targets, which have also been deployed in real world~\cite{tambe2011security}. Distinct from prior work, our problem requires choosing which centers (targets) to operate, which yields a novel mixed combinatorial and continuous optimization problem. In addition, we improve upon a prior result~\cite{yang2012computing}, that used piece wise linear approximation (PWLA) to solve a sub-sub-problem that arises in our analysis. We show that this sub-sub-problem can be computed in closed form. 

Recent work explores the complexity of quantal response players in more generality for Stackelberg games~\cite{ijcai2020-35,milec2020complexity} calling the equilibrium as Quantal Stackelberg Equilbrium (QSE). Our problem differs because of an additional combinatorial dimension (operate a subset of centers) of the defender's action space. Moreover, in contrast to the hardness results in these work, we obtain arbitrarily precise approximation for our problem in polynomial time. Thus, we identify a sub-class of games where the QSE is efficiently approximable to arbitrary precision, even with a complex defender action space.


\textit{Facility Location.} Our problem of which vaccine centers to operate might seem like the maximum capture problem in competitive facility location~\cite{Benati2002maximum,Freire2016branch,Mai2020multicut}. However, our problem is very different, as we consider security for centers and have no consideration of capturing market share and abstract away from fine-grained logistic modelling.
There are many variants of the facility location problem~\cite{procaccia2013approximate,feldman2013strategyproof,10.5555/3060621.3060641,golowich2018deep,meir2019strategyproof,ijcai2020-56,hossain2020surprising,aziz2020facility}. However, as far as we know, there is no work that considers providing \emph{security} to facilities as an optimization criteria
and most work consider \emph{only} binary decision variables. 
The interplay between discrete and continuous optimization is an important aspect of our work.

\section{Model and Problem Formulation}
We model the stated problem as a general sum Stackelberg game between a defender and a QR adversary. The defender operates a subset of centers (which changes at set frequency, e.g., weekly) and allocates security resources to operational centers. The adversary's attacks an operational center.

\textbf{Action spaces.}
The defender has \emph{a candidate set of potential centers}, denoted by $\mathcal{K}$. The variable $S$, where $S \subseteq \mathcal{K}$, denotes the  defender's choice of \emph{centers to operate}. However, not all subsets are feasible; $F(\cK) \subset 2^{\cK}$ is the feasible set of sets of centers. Suppose every center can vaccinate at least $P_{\min}$ people. Two natural restrictions are that for every $S \in F(\cK)$ to have $|S| \leq C$ and $P_{\min} |S| \geq N_P P_{\min}$ for some integer constants $C, N_P$, capturing a budget constraint and a minimum number of $N_P P_{\min}$ people to be vaccinated every round. In addition, we consider another natural constraint that the candidate center set $\cK$ is partitioned into $\cK_1, \ldots, \cK_L$ such that any feasible choice $S \in F(\cK)$ contains at least one location from each partition $l \in \{1, \ldots, L\}$ ($L < C$), that is, $|\cK_l \cap S| \geq 1$ for all $l$.
The set $\cK_l$ contains the possible operable center locations in a contiguous geographic region $l$; hence, this constraint captures fairness in the choice of allocation of vaccine centers across different geographical regions. We call this \emph{fairness in vaccine center allocation} (FVCA). 

Continuing with the defender's action description, the defender also allocates security resources to the centers $S \in F(\cK)$ that are chosen to be operational. The number of security resources is fixed and denoted by $m$. 
The defender's pure strategy is to allocate $m$ security resources to the $|S|$ operational centers. The mixed strategy is represented succinctly by $x_S = \langle x_j \rangle_{j \in S}$ ($x_S \in \cD_S^x =[0,1]^{|S|}$), which denote the marginal probability of defending the $|S|$ centers that are chosen to operate. Further, in order to provide a \emph{fairness in security allocation} (FSA), we impose the constraints: $\sum_{j \in \cK_l \cap S} x_j \leq \beta_l$ for all $l$ and some given constants $\beta_l$. In words, these constraints imposes an upper bound $\beta_l$ on the chance of protecting facilities that operate in region $l$ thereby ensuring that no geographic region is given unusual preference in terms of protection. E.g., we could have $\beta_l  \propto \frac{|\cK_l|}{|\cK|} m$ with the proportionality constant greater than $1$. 

Overall, the defender's action is $(S, x_S)$ with constraints as stated above. The adversary's pure strategy is to choose one among the $|S|$ operational centers to attack; locations in $\cK \backslash S$ have no operational center and cannot be attacked.

\textbf{Utilities.} 
For every potential center $j \in \cK$, if the center is operating and the adversary attacks $j$ and the center is protected then the defender obtains reward $r_j^d$ and the adversary obtains $l_j^a$. Conversely, if the
defender is not protecting the operating center $j$, then the defender obtains $l_j^d$ ($r_j^d > l_j^d$) and the adversary gets $r_j^a$ ($r_j^a > l_j^a$). Given
$x_j$, the expected utility of the defender and attacker for an attack on an operational center $j$ is as follows: $
U_j^d(x_j) = x_j r_j^d + (1-x_j) l_j^d$ and 
$U_j^a(x_j) = x_j l_j^a + (1-x_j) r_j^a$.
For ease of notation, we note that these utilities are linear in $x_j$ and hence we rewrite these as
\begin{align*}
U_j^d(x_j) = w_j^d x_j  + l_j^d \quad \mbox{where }  w_j^d = r_j^d - l_j^d \geq 0\\
U_j^a(x_j) = -w_j^a x_j + r_j^a \quad \mbox{where }  w_j^a = r_j^a - l_j^a \geq 0
\end{align*}
These utilities are valid for location $j \in \cK$ only when $j$ has an operational center, that is, $j \in S$. Note that $S$ is a decision variable and not fixed apriori.  



We use the QR model for the adversary's response. QR is a well-known model~\cite{mcfadden1976quantal,mckelvey1995quantal}, and used extensively in Stackelberg security games. 
The QR model is essentially a soft version of the rational best response model. 
Specifically, QR posits that the adversary will attack
an operational center $j \in S$ with probability:
\begin{align}
q_j(x_S; \lambda) = \frac{e^{\lambda(-w_j^a x_j + r_j^a)}}{\sum_{i \in S} e^{\lambda(-w_i^a x_i + r_i^a)}} \label{eq:QR}
\end{align}
Parameter
$\lambda\! \geq\!  0$ governs rationality. 
$\lambda\! =\! 0$ means least rational, as the adversary chooses its attack uniformly at
random and $\lambda\!=\! \infty$ means fully rational
(i.e., attacks a center with highest utility). Thus, QR has the flexibility to model a range of behavior. Then, defender's expected utility for $(S, x_S)$ is
$$
\cF(S,x_S) = \sum_{j \in S} q_j(x_S; \lambda) (w_j^d x_j  + l_j^d) \; .
$$
\textbf{Stackelberg equilibrium.} The Stackelberg equilibrium (also called Quantal Stackelberg Equilibrium in~\cite{ijcai2020-35}) can be computed using the following optimization:
\begin{align}
    \nonumber \max_{S \in F(\cK),x_S  \in \cD_S^x} & \cF(S,x_S)  \tag{${\sf EqOPT}$}\\
    \mbox{subject to} & \sum_{j \in S} x_j \leq m\;, \label{constraint1}\\
    & \sum_{j \in \cK_l \cap S} x_j \leq \beta_l \quad \forall l\;, \label{constraint2}
\end{align}

Constraint~(\ref{constraint1}) states that there can be at most $m$ security resources. Constraint~(\ref{constraint2}) captures the FSA fairness criteria. The $S \in F(\cK)$ in the subscript of $\max$ captures the three constraints $|S| \leq C$, $|S| \geq N_P$, and FVCA fairness criteria. The objective $\cF(S,x_S)$ is the expected utility of the defender.

\textbf{Comparison to Stackelberg security games with QR adversary.} Our model above is distinct from the standard security game with QR adversary~\cite{fang2017paws} in the aspect that we allow the defender to choose a subset of targets that then are part of the game and other targets are not part of the game. A naive exploration of all subsets is clrealy computationally infeasible, and next we present approaches to address this problem.

\section{Guaranteed Approximate Equilibrium}

As stated in the introduction, we present two approaches, one based on a MILP formulation with solution quality guarantees (in this section) and another based on a polynomial time heuristic in Section~\ref{sec:polytimeheur}. We further combine these two approaches into one hybrid approach in Section~\ref{sec:hybrid} that provides approximation guarantees for the solution, as well as similar scalability as the heuristic. 

\subsection{Common Binary Search Transformation}
We start with a transformation that is used in all our approaches. We use the Dinkelbach transform~\cite{dinkelbach1967nonlinear} to convert the fractional objective of ${\sf EqOPT}$ to a non-fractional one. We use the shorthand notation $q_j(x_S; \lambda) = N(x_j) / D(x_S)$ to express Eq.~\ref{eq:QR} succinctly (as $\lambda$ is obvious, it is not stated explicitly). By definition, $D(x_S) = \sum_{j \in S} N(x_j)$. Hence the objective of {\sf EqOPT} can be written as 
$$
\sum_{j \in S} \frac{N(x_j)}{D(x_S)} (w_j^d x_j + l_j^d).
$$
The Dinkelbach transform works by seeking to find the highest value of a threshold $\delta$ such that there exists some feasible $S,x_S$ such that $\sum_{j \in S} \frac{N(x_j)}{D(x_S)} (w_j^d x_j + l_j^d) \geq \delta$. 
The highest possible $\delta$ can be computed by a binary search where in each round of the search the feasibility problem stated above is solved for a particular $\delta_0$. For given $\delta_0$, the feasibility problem is easily solved by checking if the maximum of the following optimization is $\geq 0$ or not
\begin{align}
\max_{S \in F(\cK)} \max_{x_S \in \cD_S^x} & \sum_{j \in S} N(x_j) (w_j^d x_j\! +\! l_j^d )\! -\! \delta_0  D(x_S)  \label{prob:fraction-conversion}\tag{${\sf BOPT}$}\\
\nonumber \mbox{subject to} & \;\mbox{Constraints~(\ref{constraint1}-\ref{constraint2})}
\end{align}
\begin{algorithm}[t]
\DontPrintSemicolon
\caption{\textit{Binary Search Template}} \label{overallsolver}
$U = \max_{j \in \cK} w^d_j + l^d_j, L = \min_{j \in \cK} l^d_j, S = \phi$\;
\While{$U - L \geq \epsilon$}{
$\delta_0 =  (U+L)/2$ \;
 $obj=$ Solve \ref{prob:fraction-conversion}$(\delta_0)$ and get obj. value\;
 \lIf{$obj \geq 0$}{
 $L = \delta_0$ \llElse{
 $U = \delta_0$
 }}
 }
 return $\delta_0, S, x_S$ from the last \ref{prob:fraction-conversion} solution\;
\end{algorithm}
While the above is a common technique, for completeness, the binary search template over $\delta_0$ is shown in Algorithm~\ref{overallsolver}, where \ref{prob:fraction-conversion} is solved repeatedly (line 4) till convergence. We present different ways of solving \ref{prob:fraction-conversion} in the sequel for our different approaches. However, all our approach of solving \ref{prob:fraction-conversion} are approximate, making Algorithm~\ref{overallsolver} a binary search with inexact function evaluation, which necessitates the following result:
\begin{theorem} \label{binsearchthm}
Suppose \ref{prob:fraction-conversion} is computed with additive error of $\xi$ and runtime $T$ in line 4 of Algorithm~\ref{overallsolver}. Let $f(\delta)$ be the optimal value of \ref{prob:fraction-conversion} for $\delta$. We can show that $d|\delta - \delta'| \leq |f(\delta) - f(\delta')| \leq D|\delta - \delta'|$ for some constants $d,D > 0$ under the assumption that the absolute value of the utility parameters are bounded by constants. Moreover, Algorithm~\ref{overallsolver} runs in $O(\log(1/\epsilon) \times T)$ time with and additive error of $O(\xi + \epsilon)$.
\end{theorem}
In the rest of this paper, all our approaches will focus on approximately solving ${\sf BOPT}$ (line 4) in Algorithm~\ref{overallsolver}. The additive approximation guarantee of each approach can directly be used as $\xi$ in the above theorem to obtain the overall approximation guarantee of that approach.

\subsection{MILP Approach via Compact Linearization} 
The \ref{prob:fraction-conversion} optimization objective is separable in $x_j$'s. We exploit this and use a piecewise linear approximation (PWLA) to obtain a non-linear integer program. We follow a recipe similar to prior work~\cite{yang2012computing}, and divide the range of $x_j$, i.e., $[0,1]$, into $K$ equal intervals and represent each $x_j = \sum_{k\in [K-1]}r_{jk}$, where
$r_{jk} = 1/K$ if $k\leq \floor*{Kx_j}$ and $r_{jk} = x_j - \floor*{Kx_j}/K$ if $k = \floor*{Kx_j}+1$ and $r_{jk} = 0$ otherwise. 
However, in contrast to \cite{yang2012computing} we need additional binary variables $\theta_j$ ($\theta_j\! =\! 1$ if $j\!\in\! S$, $\theta_j\! =\! 0$ otherwise) to represent center selection, which results in \emph{bi-linear} terms of the form $\theta_jr_{jk}$ in the resultant MIP. This bilinear MIP is shown in the appendix. Bi-linear terms can be linearized using the well-known big-M  approach~\cite{Wu1997note}. But, this naive linearization results in $K|\cK|$ additional variables and $3|\cK|K$ additional constraints, making such approach not scalable at all. Next, we show that the naive bi-linear MIP can be formulated as a MILP with no additional variables and only $|\cK|$ additional constraints, with upper bounds for the approximation error given in the result below. 

\begin{theorem} \label{thm:milpapxopt}
 \ref{prob:fraction-conversion} is approximated by the following MILP
\begin{align}
    \nonumber \max_{\theta,r,z} \quad & K\sum_{j\in \cK}\theta_j(g_j(0) - \delta_0 g_j(0)) \label{prob:MILP-2}\tag{${\sf ApxOPTL}$} \\
    &\quad+ \sum_{j\in \cK} \sum_{k\in [K]} \left( \gamma^g_{jk} - \delta_0 \gamma^N_{jk}\right)  r_{jk} \nonumber \\
    \mbox{subject to} \quad& \sum_{j\in \cK}\sum_{k\in [K]} r_{jk} \leq Km\nonumber \\
    &\sum_{j\in \cK_l}\sum_{k\in [K]} r_{jk} \leq K\beta_l,\forall l\in [L] \nonumber\\
    &\theta_j \geq z_{j1},\; \forall j\in \cK\nonumber\\
    &z_{jk}\geq z_{j,k+1},\; k\in[K-1], j\in \cK \nonumber\\
    &z_{jk}/K\leq r_{jk}\leq 1/K,\; k\in [K], j\in \cK \nonumber\\
    &r_{jk}\leq z_{j,k+1}/K,\; k\in[K-1], j\in \cK \nonumber\\
    &N_P\leq  \sum_{j\in \cK}\theta_j\leq C,\; \sum_{j\in \cK_l}\theta_j\geq 1,\; l\in [L]  \nonumber\\ 
     &z_{jk}, \theta_j \in \{0,1\},\; \forall j\in \cK, k\in [K]. \nonumber
\end{align}
Let $B(S,x_S)$ be the objective function of \eqref{prob:fraction-conversion} and $S^*, x_S^*$ be an optimal solution of \eqref{prob:fraction-conversion}. Let $(\theta',z',r')$ be an optimal solution of \eqref{prob:MILP-2}, which provides solution $S',x'_S$ such that $S' = \{j|\theta'_j = 1\}$ and $(x'_S)_j = \sum_{k\in [K]}r'_{jk}$. Then, $|B(S',x'_S) - B(S^*,x^*_S)| \leq O(\frac{1}{K})$.
\end{theorem}


Using Theorem~\ref{binsearchthm} and the above result, it can be shown that when solution of \ref{prob:MILP-2} is used in line 4 of the binary search of Algorithm~\ref{overallsolver}, we attain an approximation of $O(\frac{1}{K} + \epsilon)$. However, the runtime of a MILP in the worst case is not polytime. Next, we propose a heuristic that is polytime but an approximation guarantee does not hold in rare cases.


\section{Polynomial Time Heuristic} \label{sec:polytimeheur}
Our starting point for the heuristic is problem \ref{prob:fraction-conversion}.
We aim to transform the problem further to remove constraints but the inner $\max$ problem in $x_S$ is not concave. However, a simple variable transform $y_j = e^{-\lambda w_j^a x_j}$ makes the inner problem, now in $y_S$, concave with the same optimal objective value. 
Then, we form the Lagrangian dual of the inner $\max$ problem (in $y_S$) with \emph{guaranteed same solution} due to concavity and consequent strong duality.
The optimization \emph{after Lagrangian dualizing} is 
\begin{align*}
    \nonumber & \max_{S \in F(\cK)}  \;  \min_{\nu,\mu \geq 0} \;  \max_{y_S \in \cD_S^y}\;  \phi(S, \nu, \mu,  y_S, \delta_0)   \tag{${\sf DualOPT}$}  \\
    & \mbox{where } \phi(S, \mu, \nu, y_S, \delta_0) = \\
    & \; \sum_{j \in S} N(y_j) \Big(\frac{-w_j^d \log y_j}{\lambda w_j^a}  + l_j^d \Big) - \delta_0 D(y_S) \\
    & \; - \nu \Big(\sum_{j \in S} \frac{- \log y_j}{\lambda w_j^a} - m \Big) - \sum_l \mu_l \Big(\sum_{j \in \cK_l \cap S} \frac{- \log y_j}{\lambda w_j^a} - \beta_l\Big)
\end{align*}
and $\cD_S^y$ is the Cartesian product $\bigtimes_{j \in S} [e^{-\lambda w^a_j }, 1]$, $N(y_j)\! =\! y_j e^{\lambda r^a_j}$ and $D(y_S)\! =\! \sum_{j \in S} N(y_j)$ 
and $\nu, \langle \mu_l \rangle_{l \in \{1,\ldots,L\}}$ are the dual variables. We aim to switch the outer max and min in ${\sf DualOPT}$; first, we present a general result when such switching is possible.

\subsection{A General Minimax Equality}

We present a general result about conditions for minimax equality when discrete variables are involved. We note that, as far as we know, there is no such result in literature.

\begin{theorem}\label{switchthm}
Let $f: X \times Y \rightarrow \mathbb{R}$ be a function such that the following are true:
\begin{itemize}
    \item $X$ is a set with finitely many points.
    \item $Y$ is a convex set in a Euclidean space.
    \item $f(x, \cdot)$ is continuous and convex on $Y$.
    \item For all $x \in X$ and all $b$, the sub-level sets of $f(x, \cdot)$ given by $\{y ~\vert~ f(x,y) \leq b\}$ are compact and convex (if $Y$ is compact, this condition is implied and can be removed)
    \item For any $y^* \in \arg\!\min_{y \in Y} \max_{x \in X} f(x, y)$ there is a unique $x^*$ such that $x^* =  \arg\!\max_{x \in X} f(x, y^*)$.
\end{itemize}
  Then $\min_{y \in Y} \max_{x \in X}  f(x,y) =  \max_{x \in X} \min_{y \in Y} f(x,y)$
\end{theorem}
\begin{proof}
First, compact sub-level sets of a convex function implies that the minimizer exists. This can be seen easily by choosing a $b$ such that sub-level set is nonempty and the minimizer over this sub-level set is same as global minimizer. As the sub-level set is compact, by Weierstrass extreme
value theorem the minimizer always exists. 

Then, following well-known convexity preservation transforms $\max_{x \in X} f(x,y)$ is convex. Also, $\{y ~\vert~ \max_{x \in X} f(x,y) \leq b\} = \cap_{x \in X} \{y ~\vert~ f(x,y) \leq b\}$ is compact since $\{y ~\vert~ f(x,y) \leq b\}$ is compact for each $x$. Thus, sub-level sets of the convex function $\max_{x \in X} f(x,y)$ are compact and hence a minimizer $y^*$ always exists. 

For each $x\in X$, let $y_x$ denote any element in $\arg\!\min_{y\in Y} f(x, y)$. Fix a $y^*  \in \arg\!\min_{y \in Y} \max_{x \in X} f(x, y)$. For this $y^*$ let $x^*$ be the unique value (as assumed in theorem) such that $x^*$ such that $x^* =  \arg\!\max_{x \in X} f(x, y^*)$.

We first show that
$y^* \in \arg\!\min_{y\in Y}  f(x^*, y)$ using proof by contradiction. Assume that $y^* \notin  \arg\!\min_{y\in Y} f(x^*, y)$. Our assumptions in the theorem imply that $x^*$ is unique, thus $f(x,y^*) < f(x^*, y^*)$ for all  $x\neq x^*$. 
Let 
\begin{equation}
\label{eq:th1-proof-eq1}
\delta = \min_{\substack{x \in X, x\neq x^*}} f(x^*, y^*) - f(x,y^*).    
\end{equation}
We must have $\delta>0$.
On the other hand, from the contradicting assumption and def. of $y_{x^*}$, we have that $f(x^*,y^*)> f(x^*,y_{x^*})$. 
The convexity of $f(x,y)$ in $y$ implies that, for any $\epsilon \in(0,1)$, 
\begin{align}
&(1 - \epsilon)f(x^*, y^*) + \epsilon f(x^*,y_{x^*}) \nonumber
\\ &\geq f(x^*,(1-\epsilon)y^* + \epsilon y_{x^*})\nonumber\\
&= f(x^*,y^* + \epsilon d^y),\nonumber
\end{align}
where $d^y = y_{x^*} - y^*$.
We note that, since $f(x^*,y^*)> f(x^*,y_{x^*})$ and $f$ is continuous and $y^* + \epsilon d^y$ is the line between $y^*$ and $y_{x^*}$, there is a $\epsilon_0$ such that for all $\epsilon \leq \epsilon_0$ $f(x^*,y_{x^*}) <  f(x^*, y^* + \epsilon d^y)$. Thus, plugging this into the above convexity derived result we get
\begin{equation}
\nonumber
(1 - \epsilon)f(x^*, y^*) + \epsilon f(x^*,y^* + \epsilon d^y ) > f(x^*,y^* + \epsilon d^y )
\end{equation}
which when rearranged and canceling out $(1-\epsilon)$ gives
\begin{equation}
\label{eq:th1-proof-eq3}
f(x^*,y^*) > f(x^*,y^* + \epsilon d^y )
\end{equation}
for any $\epsilon \in (0,\epsilon_0)$. Since $f(x,y)$ is continuous in $y$, we can always select $\epsilon \in (0,\epsilon_0)$ small enough such that
\begin{equation}\label{eq:th1-proof-eq2}
    |f(x,y^*) -f(x,y^* + \epsilon d^y)|< \delta/2 \mbox{ $\quad \forall x\in X$}
\end{equation}  where $\delta$ is defined in \eqref{eq:th1-proof-eq1}. 
Now, we have
\begin{align}
&f(x^*,y^* + \epsilon d^y) \stackrel{(a)}{>} f(x^*,y^*) -\delta/2  \nonumber \\ 
& \stackrel{(b)}{\geq} f(x,y^*) +\delta/2 \quad \forall x\in X \backslash \{x^*\} \nonumber \\
&\stackrel{(c)}{>} f(x,y^* + \epsilon d^y) \quad \forall x\in X\backslash \{x^*\}\nonumber
\end{align}
where $(a)$ is due to \eqref{eq:th1-proof-eq2}, $(b)$ is due to the selection of $\delta$ in \eqref{eq:th1-proof-eq1}
 and $(c)$ is due to the selection of $\epsilon$ in \eqref{eq:th1-proof-eq2}. The above set of inequalities show that $x^*$ is optimal to $\max_{x\in X} f(x,y^* + \epsilon d^y)$. Thus, from \eqref{eq:th1-proof-eq3} and the abobbe result we have
 \[
f(x^*, y^*) = \max_{x\in X} f(x,y^*) > \max_{x\in X} f(x,y^* + \epsilon d^y)
 \]
 which is contrary to  the assumption in the theorem that $y^* \in \arg\!\min_{y \in Y} \max_{x \in X} f(x, y)$. So our contradicting assumption must be false. Thus, we should have $y^* \in \arg\!\min_{y\in Y}  f(x^*, y)$. 
 
 Now we know that there is $x^*\in X$ and $y_{x^*}$ (in particular, choose $y_{x^*} = y^*$) such that $(x^*, y_{x^*})$  is optimal to $\min_{y \in Y} \max_{x \in X}  f(x,y)$. We have the following chain
 \begin{align}
     f(x^*,y_{x^*}) &=\min_{ y \in Y}\max_{x \in X} f (x,y)\nonumber\\
     &\stackrel{(d)}{\geq} \max_{x \in X} \min_{y \in Y}f (x,y)\nonumber\\
     &= \max_{x \in X} f (x,y_x) \quad \forall y_x \in \arg\!\min_{y \in Y}f (x,y)\nonumber
 \end{align}
 where $(d)$ is due to the \textit{minimax inequality}. Thus, we have $ f(x^*,y_{x^*}) = \max_{x \in X} f (x,y_x)$ since the choice of $x^*, y_{x^*}$ achieves the maximum. In other words, $(x^*,y_{x^*})$ is an optimal solution to $\max_{x \in X} \min_{y \in Y}f (x,y)$ and the minimax equality holds, i.e., 
 \[
\max_{x\in X} \min_{y \in Y} f (x,y) =  \min_{y \in Y}\max_{x \in X} f (x,y).
 \]
 We complete the proof.
 \end{proof}
\begin{figure}[t]
     \centering
     \begin{subfigure}[b]{0.23\textwidth}
         \centering \includegraphics[width=\textwidth]{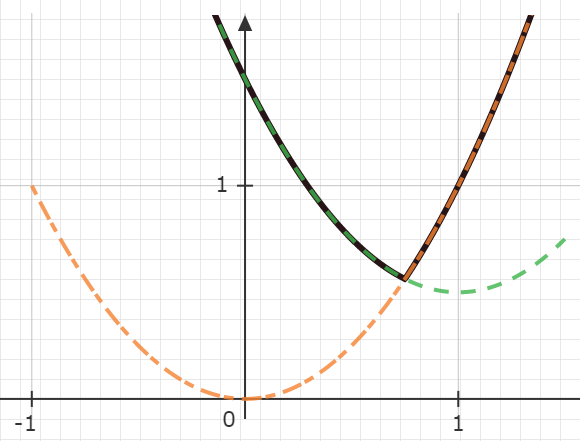}
         \caption{$f_1, f_2$, $\max(f_1, f_2)$}
         \label{fig:twoconv}
     \end{subfigure}
     \hfill
     \begin{subfigure}[b]{0.23\textwidth}
         \centering \includegraphics[width=\textwidth]{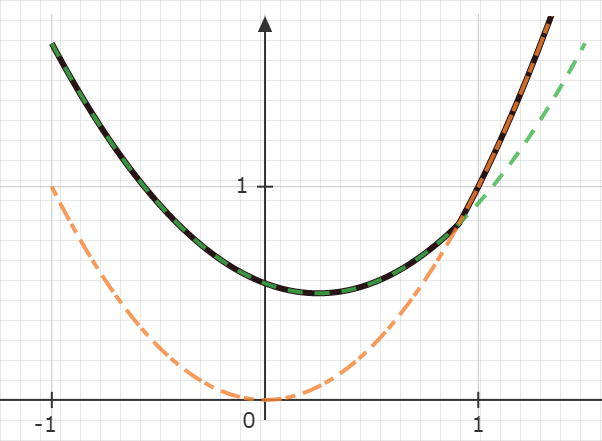}
         \caption{$f_1', f_2'$, $\max(f'_1, f'_2)$}
         \label{fig:max}
     \end{subfigure}
\caption{Simple illustration of Theorem~\ref{switchthm}. Horizontal axis is the y-axis. Max functions are in bold. $f_1, f_1'$ touch the origin.}
        \label{fig:good}
\end{figure}
We provide a simple illustration with 1d quadratic functions in Fig.~\ref{fig:good} for better intuition of the above result. The exact functions used are stated in the appendix. Fig.~\ref{fig:twoconv} shows a situation where there is a non-unique maximizer (both $1,2$) of $\min_y \max(f_1(y), f_2(y))$ for functions $f_1, f_2$ (occurring at $y=0.75$ with value $f_1(0.5)=f_2(0.5)=0.5625$). Whereas, $\max_{1,2} \min_y (f_1(y), f_2(y))$, which is min of each function separately and then the max from among those, occurs at $y=1$ with value $0.5 \neq 0.5625$. 
Fig.~\ref{fig:max} shows a situation where $\min_y \max(f_1'(y), f_2'(y))$ is uniquely determined by $f_2'$ and hence is same as $\max_{1,2} \min_y (f_1', f_2')$ .

\medskip
\subsection{A Polytime Solution for Switched Problem}
Continuing with our aim of solving ${\sf DualOPT}$, we switch the outer max and min in ${\sf DualOPT}$ to obtain ${\sf SwitchedDualOPT}$: $\min_{\nu,\mu \geq 0}\max_{S \in F(\cK)} \max_{y_S \in \cD} \phi(S,\nu, \mu, y_S, \delta_0)$. This switching is a heuristic as the uniqueness condition in Theorem~\ref{switchthm} may not hold for every possible parameter values of the problem at hand (other conditions hold); see also discussion after Lemma~\ref{closedformlemma}. 
In any case, ${\sf SwitchedDualOPT}$ provides tight bounds, which we use for a hybrid guaranteed approximate solution in the next section. 

\textbf{Solving for fixed duals.} We first show a polytime arbitrary close approximation for the inner problem in ${\sf SwitchedDualOPT}$. For any given \emph{fixed} $\nu, \mu \geq 0$, consider: 
\begin{align*}
    \nonumber\max_{S \in F(\cK)}  \;  \max_{y_S \in \cD_S^y}   & \;  \phi(S, \nu, \mu,  y_S, \delta_0) \tag{${\sf FixedDuals}$} 
\end{align*}
Observe that the objective $\phi(\cdot)$ of ${\sf FixedDuals}$ is additively separable into terms $g_j(\nu, \mu, y_j, \delta_0)$ that depend only on the single scalar variables $y_j$ as follows: 
\begin{align*}
&\phi(S, \nu, \mu,  y_S, \delta_0) = \sum_{j \in S} g_j(\nu, \mu, y_j, \delta_0) + \nu m + \sum_l \mu_l \beta_l    \\
& \mbox{ where } g_j(\nu, \mu, y_j, \delta_0) = N(y_j) \big(\frac{-w_j^d \log y_j}{\lambda w_j^a}  + l_j^d - \delta_0 \big) \\
& \qquad - 
(\nu  + \mu_l)\frac{-\log y_j}{\lambda w_j^a} \mbox{ for } l \mbox{ s.t. } j \in \cK_l 
\end{align*}
We use this separability to solve ${\sf FixedDuals}$ to any precision in polytime in Algorithm~\ref{fixeddualssolver}.
First, in lines (2-4) we maximize $g_j$ over $y_j$ for each $j \in \cK$ to obtain $h_j(\nu, \mu, \delta_0)$. With $h_j$'s, the objective of ${\sf FixedDuals}$ takes the form $\max_{S \in F(\cK)} \sum_{j \in S} h_j$ plus constants $\nu m + \sum_l \mu_l \beta_l$. Recalling the definition of $F(\cK)$, we solve this by sorting the $h_j$'s (line 5) and extracting the best $h_j$ values from each of the $L$ partitions of locations (lines 6-8). This ensures at least one center in each partition is selected. Then, among remaining locations, we choose the best (as measured by $h_j$) centers (lines 9-12) ensuring at least $N_P$ are chosen. The final choice of centers is $S_o$ (line 13).
\begin{algorithm}[t]
\caption{{\it FixedDualsSolver} ($\nu,\mu,\delta_0)$} \label{fixeddualssolver}
\DontPrintSemicolon
 $S_o = \phi, k=0$\;
 \For{$j \in \cK$}{
 Let $l$ be that index such that $j \in \cK_l$ \;
 compute $h_j(\nu, \mu,\delta_0) = \max_{y_j \in [e^{-\lambda w^a_j }, 1]} g_j( \nu, \mu,y_j, \delta_0)  $ 
 }
 In list $H$, store the indexes $j$ sorted by $h_j(\nu, \mu,\delta_0)$ in descending order.\;
 Partition $H$ into $H_1, \ldots, H_L$, such that any $j \in H_l$ satisfies $j \in \cK_l$ and each $H_l$ is still sorted by $h_j$. \;
 \For{$i \in \{1,\ldots, L\}$}{
 $S_{o} = S_{o} \cup \{H_l(0)\}$ \tcp*[f]{best $h_j$ per $\cK_l$}
\;
 }
 \While{$|S_o| < C$}{
 $j = H(k)$, $k=k+1$\;
 \If{$  h_j(\nu, \mu,\delta_0) > 0 \mbox{ or }|S_o| < N_P $}{ 
 $S_o = S_o \cup \{j \}$\;
 }
 }
 return $S^* = S_o$ and the $y^*_j$'s that maximize $h_j$
\end{algorithm}

Next, we show that there is a closed form formula for $h_j$ in line 4 of Algorithm~\ref{fixeddualssolver}. This is a concave max optimization as $g_j$ is concave in $y_j$. The closed form formula enables a much faster solving of prior security game models with QR adversary (no subset selection). The main result in~\citet{yang2012computing} was a piecewise linear approximation based optimization solution of the standard security games with QR adversary (no subset selction of targets), which we vastly improve upon by showing that the same can be computed to optimality using the closed form formula in the lemma below. In more details, we show that optimization in~\citet{yang2012computing} has a closed form \emph{convex} dual (using the lemma below) with much fewer variables (only $\mu, \nu$) than the primal; hence the whole expensive piecewise linear approximation approach in~\cite{yang2012computing} can just be replaced by an easy gradient descent on the dual program. This result provides scalability in all the other works in wildlife security games~\cite{fang2017paws} that use the result from~\citet{yang2012computing}.
\begin{lemma} \label{closedformlemma}
The solution of the problem in line 4 of Algorithm~\ref{fixeddualssolver} is
$$
y^*_j =  e^{-\lambda w^a_j \max(0,\min(\beta_j,1))}
\mbox{ where $\beta_j$ is}$$ 
$$
\frac{1}{\lambda w^a_j }\Big[ 1- \frac{\lambda w^a_j}{w^d_j}(l_j^d - \delta_0) - W\big(\frac{\nu  + \mu_l}{w^d_j} e^{1-\lambda r^a_j - \frac{\lambda w^a_j}{w^d_j}(l_j^d - \delta_0)}\big) \Big]
$$
and $W$ is the Lambert W function~\cite{corless1996lambertw}.
\end{lemma}

For further notational ease, let $\Phi(S, \nu, \mu,  \delta_0) =\max_{y_S \in \cD_S^y}\phi(S, \nu, \mu, y_S, \delta_0)$. Using the solution $y^*_S$ from the above lemma, we get that $\Phi(S, \nu, \mu, \delta_0) = \phi(S, \nu, \mu, y^*_S, \delta_0)$. Note that $y^*_S$ is still a function of $\nu, \mu, \delta_0$. It can be readily checked that $\Phi$ satisfies the conditions of Theorem~\ref{switchthm} with $\Phi$ as $f$, and $\nu, \mu$ as $y$, and $S$ as $x$ ($\delta_0$ is a constant), \emph{except} for the uniqueness of $S^*$ ($x^*$ in Theorem~\ref{switchthm}) for optimal $\nu^*, \mu^*$ ($y^*$ in Theorem~\ref{switchthm}).
However, observe that in Algorithm~\ref{fixeddualssolver}, the solution $S^*$ (for given $\nu^*, \mu^*$) may be non-unique only if $h_j\! =\! h_i$ for some $i\! \neq\! j$; the stringent equality needed makes non-uniqueness highly unlikely; indeed, we encounter non-uniqueness in only 1.1\% cases in experiments. We prove the following:
\begin{lemma} \label{fixeddualssolverproperty}
\begin{enumerate}
\item Algorithm~\ref{fixeddualssolver} runs in poly time $O(|\cK| \log |\cK|)$.
\item Algorithm~\ref{fixeddualssolver} solves ${\sf FixedDuals}$ to any arbitrary fixed precision. 
\end{enumerate}
\end{lemma}

\textbf{Gradient descent on duals.}
Next, in order to solve ${\sf SwitchedDualOPT}$, we  perform projected gradient descent (PGD) on dual variables $\nu, \mu$. Recall that the inner problem in ${\sf SwitchedDualOPT}$ is  $\max_{S \in F(\cK)} \max_{y_S \in \cD_S^y} \phi(S, \nu, \mu, y_S, \delta_0)$, which, with slight abuse of notation, we refer to as ${\sf FixedDuals}(\nu, \mu, \delta_0)$, which we already solved in Algorithm~\ref{fixeddualssolver} for given $\nu, \mu, \delta_0$. The use of PGD is justified by:
\begin{proposition} \label{prop:convexity}
${\sf FixedDuals}(\nu, \mu, \delta_0)$ is convex in $\nu, \mu$.
\end{proposition}
\begin{proof}
We again skip writing $\delta_0$ for ease of notation. Using the notation introduced and Lagrangian duality, the inner $\Phi(S,\nu, \mu) = \max_{y_S \in D} \phi(S,\nu, \mu,y_S)$ problem is convex in $\nu, \mu$. The function in the theorem is $\max_{S \in F(\cK)} \Phi(S, \nu, \mu)$. As this is a max over multiple convex functions $\Phi(S, \nu, \mu)_{S \in F(\cK)}$, this function is convex.
\end{proof}
Combining all sub-results for the polynomial heuristic, in Algorithm~\ref{SwitchedDualOptSolver}
PGD is used in lines 3-4, where $\mathbb{P}_{\nu,\mu\geq 0}$ denotes projection to the space $\nu, \mu \geq 0$. The gradient of ${\sf FixedDuals}$ w.r.t. $\nu,\mu$ can be computed using Danskin's Theorem~\cite{bertsekas1998nonlinear} (details in appendix). 
Algorithm~\ref{alg:vacsec}, when plugged in as solver for \ref{prob:fraction-conversion} in line 4 of the binary search Algorithm~\ref{overallsolver} is the full polynomial time heuristic. 

\begin{algorithm}[t]
\DontPrintSemicolon
\caption{\textit{SwitchedDualOptSolver}$(
\xi, \delta_0)$} \label{SwitchedDualOptSolver} \label{alg:vacsec}
$(\nu^0, \mu^0) = 0$ \;
\Repeat{objective changes by less than $\xi$}{
$(\nu^t, \mu^t) = \mathbb{P}_{\nu\mu \geq 0}\big((\nu^{t-1}, \mu^{t-1}) - \eta_t \nabla_{\nu,\mu} {\sf FixedDuals}(\nu^{t-1},\mu^{t-1}, \delta_0) \big) $\;
$S, y_S = {\it FixedDualsSolver}(\nu^t, \mu^t, \delta_0)$\;
 } 
 
 return $S, y_S$
\end{algorithm}

Using Theorem~\ref{binsearchthm} and Lemma~\ref{fixeddualssolverproperty}, it can be readily seen that the heuristic has a runtime of $O(\log(\frac{1}{\epsilon})\frac{|\cK| \log |\cK|} { \xi})$ (gradient descent takes $O(\frac{1} { \xi})$ iterations under mild smoothness condition). Also, if the uniqueness condition of Theorem~\ref{switchthm} holds for any problem instance then we would obtain $O(\xi + \epsilon)$ approximation. However, as stated earlier, the uniqueness might not hold in rare cases and hence we next propose a hybrid approach combining the advantages of this heuristic and the earlier MILP approach.

\section{Hybrid Approach}\label{sec:hybrid}
This hybrid approach is inspired by the observation that the heuristic is time efficient and if the minimax equality hold, the resulting solution is optimal for ${\sf EqOPT}$. If the minimax eaulity does not, we can still use solution of ${\sf SwitchedDualOPT}$ to construct tight lower and upper bounds for the MILP approach with guaranteed solution quality. Towards that end, we make use of the following result:
\begin{theorem}
\label{th:hybrid}
If we run the heuristic  (Algorithm~\ref{SwitchedDualOptSolver} plugged in line 4 of Algorithm~\ref{overallsolver}) and obtain solution $(\overline{\delta}_0,\overline{S},\overline{x}_S)$, then with $S^*, x^*_S$ optimal for ${\sf EqOPT}$ there exists a small enough $ \xi$ (in Algorithm~\ref{SwitchedDualOptSolver}) such that
(1) $|\cF(S^*,x^*_S) - \cF(\overline{S},\overline{x}_S)| \leq |\overline{\delta}_0 - \cF(\overline{S},\overline{x}_S)| + 2\epsilon$ and (2)
$\cF(\overline{S},\overline{x}_S)$ and $\overline{\delta}_0+2\epsilon$ can be used as a lower and upper bounds for the MILP approach (${\sf ApxOPTL}$ plugged in line 4 of Algorithm~\ref{overallsolver}).
\end{theorem}

Theorem~\ref{th:hybrid} implies that if $\overline{\delta}_0$ is sufficiently close to $\cF(\overline{S},\overline{x}_S)$, then $(\overline{S},\overline{x}_S)$ is a near-optimal solution to ${\sf EqOPT}$. 
Using the above result, we design a hybrid approach combining the MILP and the heuristic to efficiently find a near-optimal solution in Algorithm~\ref{alg:hybrid}. 
\begin{algorithm}[t]
\DontPrintSemicolon
\caption{\textit{Hybrid algorithm}} \label{alg:hybrid}
run heuristic (Algorithm~\ref{SwitchedDualOptSolver} used in Algorithm~\ref{overallsolver}) to get $(\overline{\delta}_0,\overline{S},\overline{x}_S)$\\
\eIf{$|\overline{\delta}_0 -\cF(\overline{S},\overline{x}_S)|\leq \epsilon$}{ 
 return $(\overline{S},\overline{x}_S)$\;
 }
 {
 run MILP approach (Algorithm~\ref{overallsolver} with ${\sf ApxOPTL}$), with $L = \cF(\overline{S},\overline{x}_S)$, $U = \overline{\delta}_0+2\epsilon$, obtain $(S^*,x_S^*)$\;
 run heuristic with fixed $S=S^*$ to improve $x^*_S$\;
 return the last solution obtained.
 }
\end{algorithm}
{It can be seen from the above result that if the algorithm stops at line 3, then the returned solution is additive $3\epsilon$-optimal to ${\sf EqOPT}$, i.e., $\cF(S^*,x^*_S) - \cF(\overline{S},\overline{x}_S)\leq 3\epsilon$. Otherwise, if the algorithm stops at line 7, then a guarantee is already established via Theorem~\ref{thm:milpapxopt}, i.e., $\cF(S^*,x^*_S) - \cF(\overline{S},\overline{x}_S) \leq O(1/K + \epsilon)$. So, it is guaranteed that  a  solution returned by Algorithm~\ref{alg:hybrid} is additive $O(1/K + \epsilon)$-optimal to  ${\sf EqOPT}$. In line 6, we run the heuristic again to further improve the solution of the MILP.
In experiments we show that in most of the cases, the hybrid algorithm stops at line 3, making it much faster than the MILP approach.}

\section{Experiments}
Our experiments are simulated over 10 random
instances for each measurement that we report. In each game instance, payoffs are chosen
uniformly randomly, from 1 to 10 for $r^d_j$ and $r^a_j$ and from -10 to -1 for $l^d_j$ and $l^a_j$. Following past user studies~\cite{yang2011improving}, the
parameter $\lambda$ of the QR model is chosen as $0.76$.  We select $C = \left \lfloor{2\cK/3}\right \rfloor$, 
$N_P = \left \lfloor{\cK/2}\right \rfloor$,
$m = |\cK|/10$, and split the set of centers into $L\! =\! 5$ disjoint partitions of equal size. For each partition $l$, we choose $\beta_l = 2m/L$.
All experiments were conducted using Matlab on a Windows 10 PC with Intel i7-9700
CPUs (3.00GHz).

We compare our algorithm (Alg~\ref{alg:hybrid}, denoted as \textit{Hybrid}) with two baseline methods. One is based on the convex optimization approach  (denoted as \textit{ConvexOpt}) proposed in prior work~\cite{yang2012computing}; this method is not capable of choosing which center to operate hence we run it with all the centers chosen as operational. The other method is based on a two-steps procedure (denoted as \textit{TwoSteps}); for this we first solve ${\sf EqOPT}$ with fixed $S = \cK$ (using \textit{ConvexOpt}) to find a strategy $x^*$. We then use this strategy 
to select at least $N_P$ and at most $C$ centers from $\cK$ 
by sorting the individual rewards $\{w^d_j x^*_j + l^d_j, j\in \cK\}$ and selecting
$N_P$ centers with highest rewards and  then, from the remaining centers, selecting no more than $C\!-\!N_P$ centers with highest and positive rewards. Then, a subset $S^*$ is selected and we solve ${\sf EqOPT}$ with fixed $S\! =\! S^*$ to re-optimize the strategy. 
 
\begin{figure}[t] 
\centering
    \includegraphics[width=0.50\linewidth]{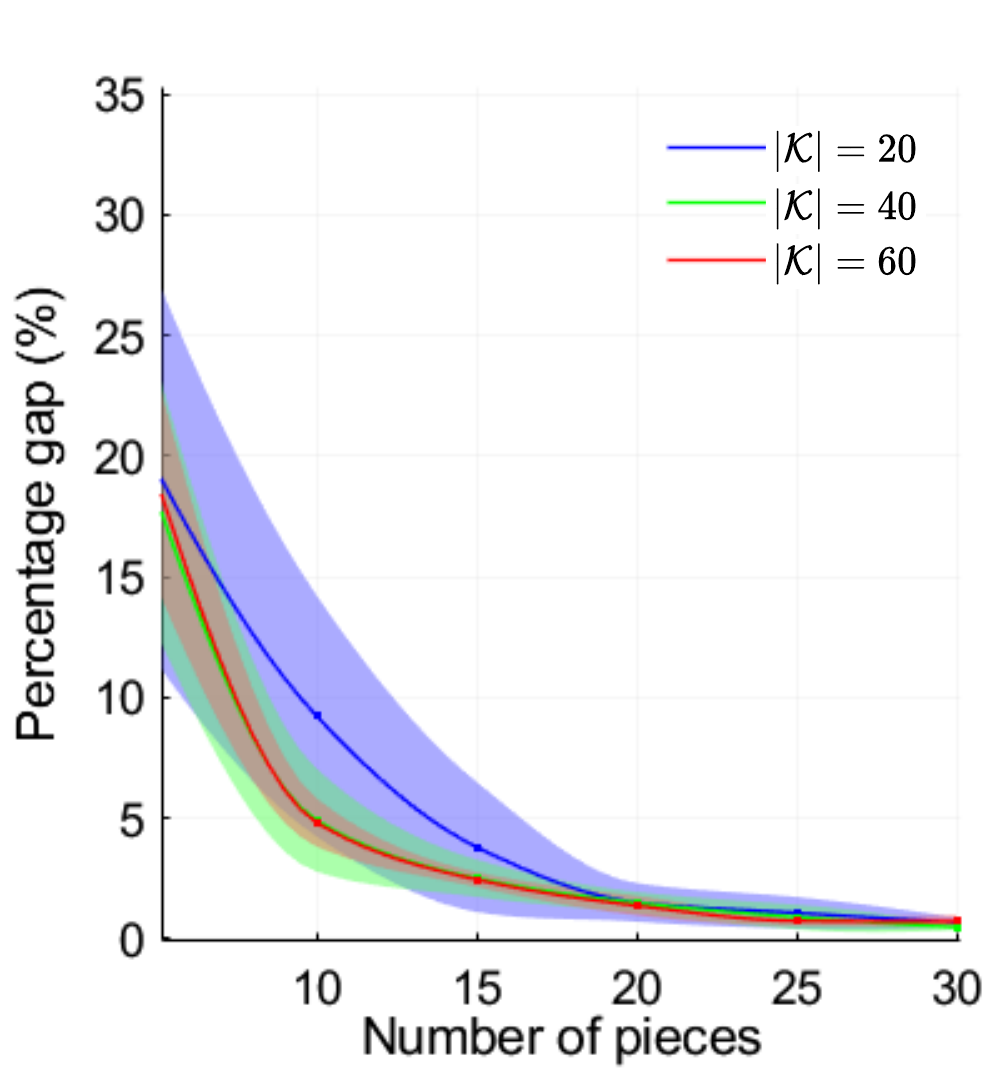} 
    \caption{Performance of the PWLA as  $K$ increases.} 
    \label{fig:milp-convergence} 
\end{figure}

We first determine the number of pieces $K$ needed in our PWLA approach for a good approximation, noting that, as is typical for approximation algorithms, the bound in Theorem~\ref{thm:milpapxopt} can be too conservative for average case problems. We vary $K$ in $\{5,10,\ldots,30\}$ and compute the percentage gap between the objective values given by PWLA with $K$ pieces and with a large $\overline{K} = 200$ pieces.
Figure~\ref{fig:milp-convergence} plots the percentage gaps for $|\cK| \in\{20,40,60\}$ and we see that the percentage gaps become relatively small (less than $1.4\%$) if $K\geq 20$.
Thus, we fix $K=20$ for the rest of the experiments.
 
\begin{figure}[t] 
\centering
    \includegraphics[width=1.0\linewidth]{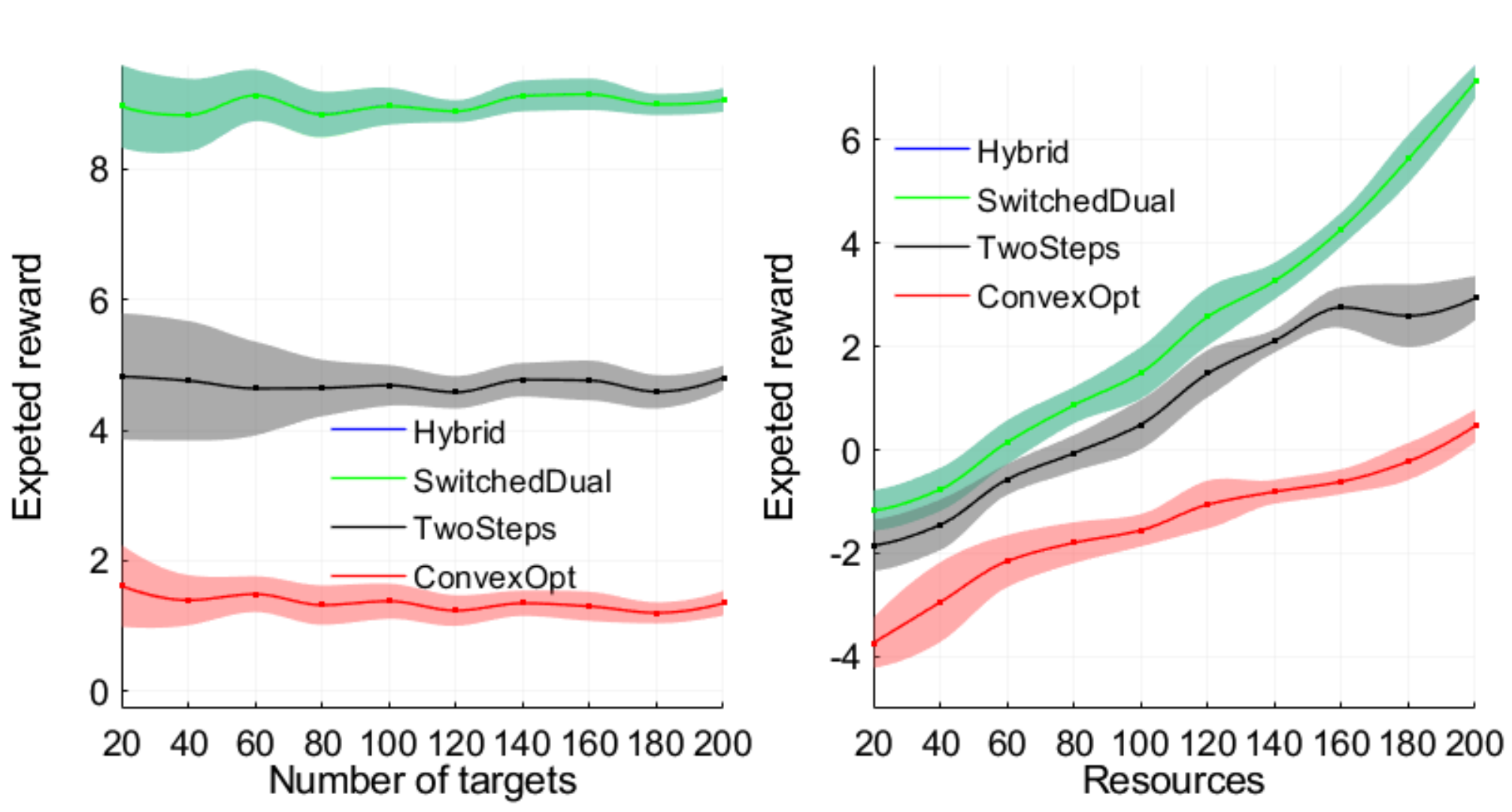} 
    \caption{ Expected reward comparison, the curves given by \textit{Hybrid} and \textit{SwitchedDual} are almost identical.} 
    \label{fig:expected-reward} 
\end{figure}

\textbf{Expected rewards comparison.}
 The means and standard errors of the  expected rewards of different approaches are plotted in Fig.~\ref{fig:expected-reward}, where in the left figure we vary the number of centers from 20 to 200 and set resource budget as $m=|\cK|/10$, and in the right figure we fix $|\cK| = 50$ and vary the resource budget $m$ from 2 to 20.
 The expected rewards of \textit{SwitchedDual} are equal to those of \textit{Hybrid} for 178/180 test instances, and only slightly smaller for 2/180 instances. This shows that the \textit{minimax equality} holds with high probability.  

Also, \textit{Hybrid} and \textit{SwitchedDual} consistently outperform other methods.
In particular, for varying number of centers, \textit{Hybrid} provides 85\%-96\% larger rewards than \textit{TwoSteps}. For varying resource budget, the improvements are up to 142\%. \textit{ConvexOpt} returns very low rewards, revealing clear benefit of selecting centers instead of operating all possible centers.

\begin{figure}[t] 
\centering
    \includegraphics[width=1.0\linewidth]{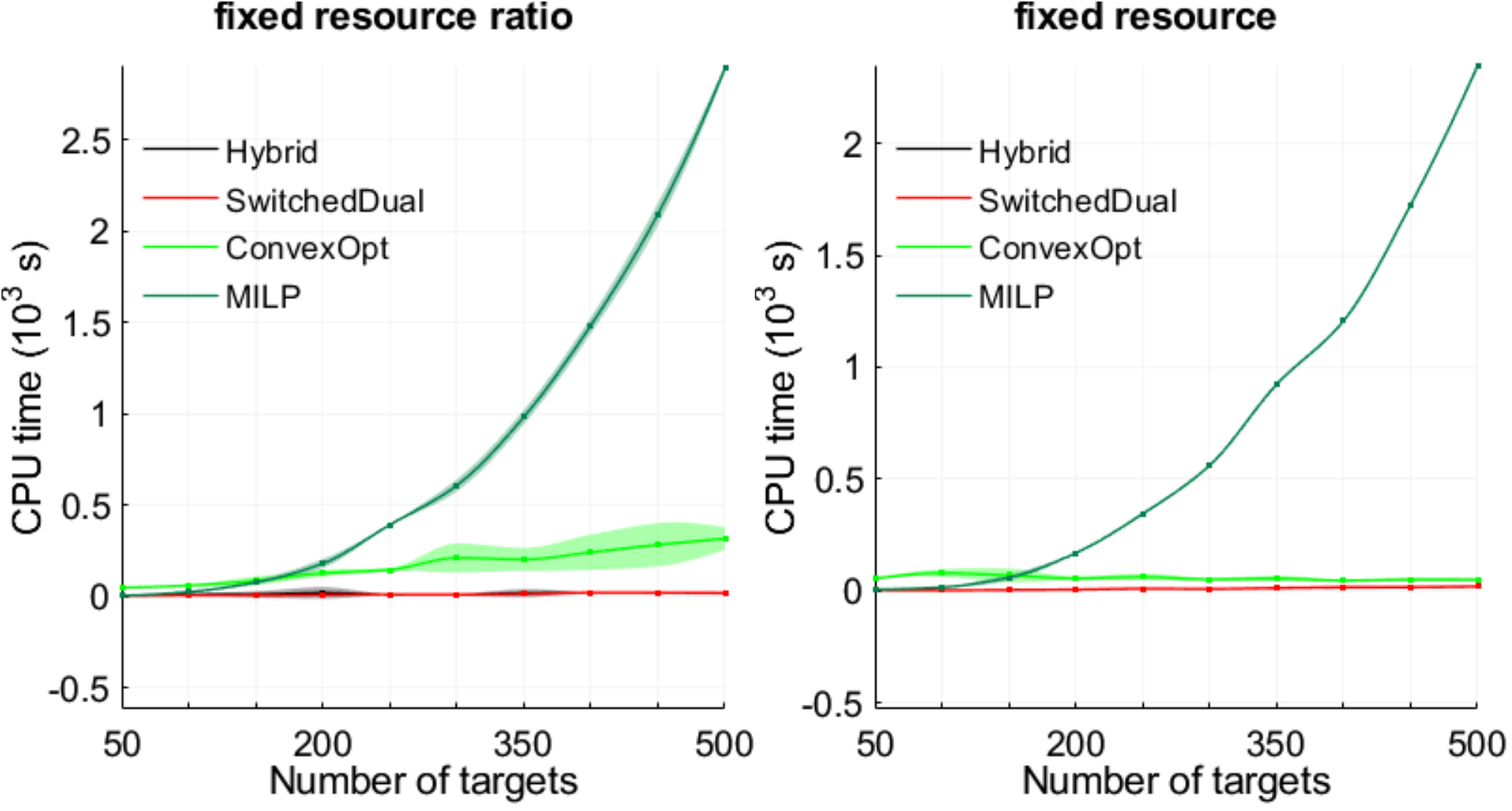} 
    \caption{Computational scalability comparison, the curves given by \textit{Hybrid} and \textit{SwitchedDual} are almost identical.} 
    \label{fig:scalability} 
\end{figure}

\textbf{Computational scalability.}
In this experiment, we run  \textit{Hybrid}, 
\textit{SwitchedDual}, \textit{ConvexOpt} and {\textit{MILP}}, where MILP refers to using ${\sf ApxOPTL}$. 
We vary the number of potential centers from 50 to 500 with two settings, one with a fixed resource ratio as $m = 0.1|\cK|$ and one with a fixed resource budget $m = 20$. Fig.~\ref{fig:scalability} shows the means and standard errors of the CPU time of the four approaches over 10 repetitions.
We also see that the curves given by the \textit{Hybrid} and \textit{SwitchedDual} are almost identical, indicating that Algorithm~\ref{alg:hybrid} mostly stops at line 3, i.e., the minimax equality holds. There are only a few instances of $|\cK| = 200$ or $|\cK| = 350$ that Algorithm~\ref{alg:hybrid} stopped at line 7, noting that even in these cases, the total time is still about 5 to 10 times less than the time required by the \textit{MILP}, due to the tightness of the bounds provided by the \textit{SwitchedDual} in line 5. Note that \textit{ConvexOpt} solves an easier problem with no selection of centers, but even then \textit{Hybrid} runs faster than \textit{ConvexOpt} on average.
 
The time taken by \textit{MILP} grows very fast as the number of centers increases whereas the time required by \textit{Hybrid} and \textit{SwitchedDual} is stable and small. More precisely, for fixed resource ratio, with $|\cK| = 500$, the times required by \textit{Hybrid, SwitchedDual}, \textit{ConvexOpt}  and \textit{MILP} are about 4, 4, 400 and 4700 (secs), respectively. 
To further demonstrate scalability, we increased the number of centers to $5000$ and \textit{Hybrid} and \textit{SwitchedDual} finished in 70 seconds.
In summary, the results show the superiority of our algoriTheorem 

We evaluate the impact of the FSA constraints on the fairness of security resource distribution. To this end, we select $|\cK| = 20$  and also divide all the centers into $L=5$ partitions of equal size. 
To better illustrate the fairness, we add 5 units to $r^a_j$, $\forall j\in\cK_1$. This implies that the adversary will get more rewards if attacking a target in Partition \#1. We also tighten the selection of parameters $\beta_l$ by choosing $\beta_l = 1.2m/L$, for all $l=1,\ldots,L$.

\begin{figure}[t] 
\centering
    \includegraphics[width=1\linewidth]{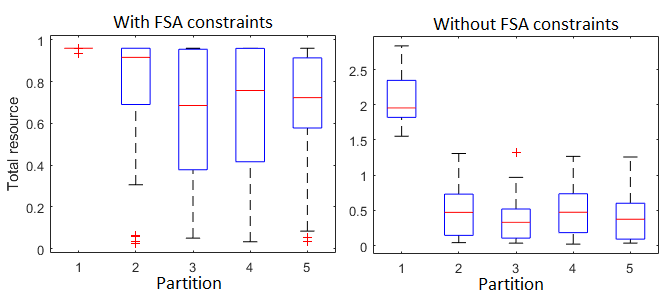} 
    \caption{Fairness in security allocation over vaccine centers} 
    \label{fig:fairness} 
\end{figure}

The box plot in Figure~\ref{fig:fairness} reports the distributions of the total resources assigned to the 5 partitions, with and without the FSA constraints. Without the FSA constraints, the first partition gets a high chance of being protected (allocating an expected number of more than two resources to the four centers in Partition \#1), thus lowering the protection of other partitions. On the other hand, the FSA constraints maintains a fairness of security allocation between partitions. 

\medskip

In short, Fig.~\ref{fig:fairness} clearly shows us the role of the FSA constraints in maintaining a fairness in security allocation. 

\section{Conclusion}
We proposed a model for security of vaccine delivery in underdeveloped and security risk-prone areas and presented an efficient solver using a novel strong duality result with discrete variables. As such, a number of other schemes require security (setting up voting centers, medical camps, etc.). We believe our model can serve as a basis for formally modeling such problems and its variants. Our technical approach can also inform other techniques where a mix of discrete and continuous variables are used. 







\bibliographystyle{ACM-Reference-Format} 
\bibliography{main}

\appendix
\section{Details of Omitted Parts and Additional Experiments}


\subsection{Bilinear ILP and MILP} \label{sec:ilp}

For ease of notation,  let $g_j(x_j) = N(x_j)(w^d_jx_j +l^d_j)$. We divide the range of $x_j$, i.e., [0,1], into $K$ equal intervals and represent each $x_j = \sum_{k\in [K-1]}r_{jk}$, where
$r_{jk} = 1/K$ if $k\leq \floor*{Kx_j}$ and $r_{jk} = x_j - \floor*{Kx_j}/K$ if $k = \floor*{Kx_j}+1$ and $r_{jk} = 0$ otherwise. 
We then approximate
the univariate functions $g_j(x_j)$ and $N(x_j)$ as 
\begin{align}
    N(x_j) \approx N_j(0)+ {1}/{K}\sum_{k\in [K]} \gamma_{jk}^N r_{jk} \nonumber \\
    g_j(x_j) \approx g_j(0)+ {1}/{K}\sum_{k\in [K]} \gamma^g_{jk} r_{jk} \nonumber
\end{align}
where $N_j(0) = e^{\lambda r^a_j}$,  and $\gamma_{jk}^N, \gamma^g_{jk}$
are the slopes of $N_j(x_j), g_j(x_j)$ in each small interval in $[0,1]$, defined as
\begin{align}
\gamma_{jk}^N &= K\left(N_j(k/K) - N_j((k-1)/K) \right), \; k\in [K] \nonumber \\
\gamma_{jk}^g &= K\left(g_j(k/K) - g_j((k-1)/K) \right), \; k\in [K] \nonumber
\end{align}
To include $r_{jk}$ into the optimization model, we introduce binary variables $z_{jk}\in\{0,1\}$ such that $z_{jk}\geq z_{j,k+1}$ for $k\in[K-1]$. We constrain $0\leq r_{jk}\leq 1/K$ for all $k$, and use $r_{jk} \geq z_{jk}/K$ to force $r_{jk}$ to $1/K$ if $z_{jk} = 1$, and use $r_{jk} \leq z_{j,k+1}/K$ to force $r_{jk} = 0$ if $z_{j,k+1} = 0$. For the first $k$ s.t. $z_{jk} =0$ (representing $k= \floor*{x_jK}+1$) the only applicable constraint is $0\leq r_{jk}\leq 1/K$. Moreover, we also include binary variables $\theta_j \in \{0,1\}$, $j\in \cK$ to represent a subset $S\subset \cK$, i.e., $\theta_j = 1$ if $j\in S$ and $\theta_j = 0$ otherwise. Combining all these, we can approximate \eqref{prob:fraction-conversion} by the following binary \textit{nonlinear} program
\begin{align}
    \nonumber \max_{\theta,r,z} \quad & K\sum_{j\in \cK}\theta_j(g_j(0) - \delta_0 N_j(0)) \label{prob:MILP-1}\tag{${\sf ApxOPT}$} \\
    &\quad+ \sum_{j\in \cK} \sum_{k\in [K]} \left( \gamma^g_{jk} - \delta_0 \gamma^N_{jk}\right) \theta_j r_{jk} \nonumber \\
    \mbox{subject to} \quad& \sum_{j\in \cK}\sum_{k\in [K]} \theta_jr_{jk} \leq Km \\
    & \sum_{j\in \cK_l}\sum_{k\in [K]} \theta_jr_{jk} \leq K\beta_l,\;\forall l\in [L] \\
    &z_{jk}\geq z_{j,k+1},\; k\in[K-1], j\in \cK \label{eq:firstref}\\
    &z_{jk}/K\leq r_{jk}\leq 1/K,\; k\in [K], j\in \cK \\
    &r_{jk}\leq z_{j,k+1}/K,\; k\in[K-1], j\in \cK \\
    & N_P\leq \sum_{j\in \cK}\theta_j\leq C,\; \sum_{j\in \cK_l}\theta_j\geq 1,\; l\in [L]  \\ 
     &z_{jk}, \theta_j \in \{0,1\},\; \forall j\in \cK, k\in [K] \label{eq:lastref}
\end{align}


The above program contains bi-linear terms $\theta_jr_{jk}$, which can be linearized using the well-known ``big-M''  approach~\cite{Wu1997note}. However, this requires $K|\cK|$ additional variables and $3|\cK|K$ additional constraints, making such approach not scalable at all. In the following result we show that  \eqref{prob:MILP-1} can be formulated as a MILP with no additional variables and only $|\cK|$ additional constraints. 
In particular, we claim that \eqref{prob:MILP-1} is equivalent to the following MILP
\begin{align}
    \nonumber \max_{\theta,r,z} \quad & K\sum_{j\in \cK}\theta_j(g_j(0) - \delta_0 g_j(0)) \tag{${\sf ApxOPTL}$} \\
    &\quad+ \sum_{j\in \cK} \sum_{k\in [K]} \left( \gamma^g_{jk} - \delta_0 \gamma^N_{jk}\right)  r_{jk} \nonumber \\
    \mbox{subject to} \quad& \sum_{j\in \cK}\sum_{k\in [K]} r_{jk} \leq Km \nonumber\\
    & \sum_{j\in \cK_l}\sum_{k\in [K]} r_{jk} \leq K\beta_l,\;\forall l\in [L] \nonumber\\
    &\theta_j \geq z_{j1},\; \forall j\in \cK\nonumber\\
     & \mbox{and constraints~(\ref{eq:firstref}-\ref{eq:lastref})} \nonumber
\end{align}
The above MILP is exactly the same one as shown in the statement of  Theorem~\ref{thm:milpapxopt} in the main paper.
We show the proof of the above claim in the proof of Theorem~\ref{thm:milpapxopt}.

\subsection{Variable Transformation}
Using $y_j = e^{-\lambda w_j^a x_j}$, the complete transformed optimization is
\begin{align}
    \nonumber \max_{S \in F(\cK)} \max_{y_S \in \cD} & \sum_{j \in S} N(y_j) \Big(\frac{-w_j^d \log y_j}{\lambda w_j^a}  + l_j^d \Big) - \delta_0 D(y_S) \\
    \mbox{subject to} & \sum_{j \in S}  \frac{- \log y_j}{\lambda w_j^a} \leq m\;, \label{constraint5}\\
    & \sum_{j \in \cK_l \cap S}  \frac{- \log y_j}{\lambda w_j^a} \leq \beta_l \quad \forall l\;, \label{constraint6}
\end{align}
where $\cD$ is the convex set given by the Cartesian product $\bigtimes_{j \in S} [e^{-\lambda w^a_j }, 1]$, $N(y_j) = y_j e^{\lambda r^a_j}$ and $D(y_S) = \sum_{j \in S} N(y_j)$. 
\subsection{Gradient using Danskin's Theorem}
Briefly, Danskin's Theorem states that given $f(x) = \max_{z \in Z} g(x,z)$ for compact set $Z$ and $g(x,z)$ convex in $x$ for all $z$, and a unique maximizer $z^*$ of $\max_{z \in Z} g(x,z)$ for any $x$, then we have
$$
\frac{\partial f}{\partial x} = \frac{\partial g(x, z^*)}{\partial x} \mbox{ where } z^* \mbox{ is treated as a constant}.
$$
Recall that ${\sf FixedDuals}$ is $\max_{S, y_S \in D} \phi(S,\nu, \mu,y_S)$. By Danskin's theorem, given unique optimal $S^*, y_S^*$ (which we assume in this heuristic) and $\phi$ convex in $\nu,\mu$, we obtain the required derivative.
an be obtained as 
$$
\frac{\partial \phi(S^*,\nu, \mu,y_S^*)}{\partial \nu} = \sum_{j \in S^*}\frac{\log y_j^*}{\lambda w_j^a} + m, $$

$$\frac{\partial \phi(S^*,\nu, \mu,y_S^*)}{\partial \mu_l} = \sum_{j \in \cK_l \cap S^*}\frac{\log y_j^*}{\lambda w_j^a} + \beta_l \quad \forall l
$$

\subsection{Functions used in Fig.~\ref{fig:good}}

The two functions in Fig.~\ref{fig:twoconv} are $f_1(y) = y^2$ and $f_2(y) = 0.5 + (y-1)^2$.
$\max_{1,2} \min_x (f_1(x), f_2(x)) $ considers a min of each function separately and then chooses the max from among those. $f_1$ has min value $0$ at $x=0$ and $f_2$ has min value $0.5$ at $x=1$. Thus, the outer max chooses solution value $0.5$ at $x=1$. 
The solution of $\min_x \max_{1,2}  (f_1(x), f_2(x)) $ is at $x = 0.75$ with value $0.5625$, which is a different solution from the max-min case. 
Also, note that at $x = 0.75$, we have $f_1(0.75) = f_2(0.75)$ thus, the inner problem $\max_{1,2}  (f_1(x), f_2(x))$ does not have a unique solution. This shows that uniqueness is needed for the minimax equality to hold.

The two functions in Fig.~\ref{fig:max} are $f_1'(y) = y^2$ and $f_2'(y) = 0.5 + 0.75(y-0.25)^2$. The min of the min max occurs at $y= 0.25$ and is uniquely determined by $f_2'$.


\section{Missing Proofs}

\subsection{Proof of Theorem~\ref{binsearchthm}}
\begin{proof}
We first prove a claim that $f$ is a strictly monotonic decreasing function of $\delta$ and hence $f^{-1}$ is well-defined. This can be verified readily as $D(x_S) > 0$ for any feasible $x_S$; suppose optimal solution value is $f({\delta})$ for $\delta$. For contradiction, assume $f({\delta'}) \geq f({\delta})$ for some $\delta' > \delta$ with optimal solution $S', x'_S$ for $\delta'$. Clearly, $S', x'_S$ is feasible with $\delta$ also and achieves objective value $\sum_{j \in S'} N(x'_j) (w_j^d x'_j  + l_j^d ) - \delta D(x'_S) = f({\delta'}) + (\delta'-\delta) D(x'_S)$ which is more than $f(\delta)$, a contradiction. Thus, we must have $f({\delta'}) < f({\delta})$. Hence, $f^{-1}$ is well-defined. Also, the Lipschitz constants are derived at the end.

Next, as defined, $f(\delta)$ is the ideal value of $obj$ in the algorithm if the \ref{prob:fraction-conversion} solver has no error. Suppose $S_{\delta}, x_{S, \delta}$ is the optimal for some $\delta$ and approximation provides $S',x'_S$ with objective value within $\xi$ of the objective value with $S_{\delta}, x_{S, \delta}$. The $\xi$ approximation guarantee gives for any $\delta$
\begin{align}
\nonumber & |f(\delta) - {\sf BOPT}(\delta, S', x'_S)| = \\ 
& |{\sf BOPT}(\delta, S_{\delta}, x_{S, \delta}) - {\sf BOPT}(\delta, S', x'_S)| \leq \xi \label{eq:xierror}
\end{align}
Let $\delta^*$ be a fixed value such that $f(\delta^*)=0$. We seek $\delta^*$ in this algoriTheorem By definition, $\delta^* = \cF(S^*, x_S^*)$, where $S^*, x_S^*$ is an optimal variable value for ${\sf EqOPT}$. By Lipschitzness assumption in the theorem, for any $\delta$
\begin{align}
|\delta -\delta^*| \leq (1/d)|f(\delta) - f(\delta^*)| = (1/d)|f(\delta)| \label{eq:KLip}
\end{align}

When the binary search stops then $U-L \leq \epsilon$. Suppose $U$ and $L$ was last updated with corresponding $\delta_U$ and $\delta_L$ (in line 5). 
Then, $\delta_U - \delta_L \leq \epsilon$. Also, suppose when the binary search stops the solution for the  variables are $\overline{S}, \overline{x}_S$. We hope to have $\delta^* \in (\delta_L, \delta_U)$ but cannot claim this for sure because of the $\xi$ error.

We consider the case that $\delta^*$ does not lie in $(\delta_L, \delta_U)$ and we bound how far $\delta^*$ is from $\delta_U, \delta_L$. WLOG, we assume $U$ was last updated before binary search stopped (the case of $L$ last updated can be handled exactly analogously). Next, suppose the output of binary search is $\overline{\delta}_0, \overline{S}, \overline{x}_S$. We know that $\overline{\delta}_0 = \delta_U$. We know that objective ${\sf BOPT}(\delta_U, \overline{S}, \overline{x}_S)$ with $\delta_U$ was negative (that is, $obj$ was negative) because $U$  was updated. 

Suppose $\delta^* > \delta_U$, thus, due to decreasing $f$ we obtain $f(\delta^*) < f(\delta_U)$ meaning $f(\delta_U)$ is positive.  Using Equation~\ref{eq:xierror}, we get
$$
|f(\delta_U) - {\sf BOPT}(\delta_U, \overline{S}, \overline{x}_S) | \leq \xi
$$
With the knowledge that ${\sf BOPT}(\delta_U, \overline{S}, \overline{x}_S)$ is negative and $f(\delta_U)$ is positive, we can readily infer that $f(\delta_U) \leq \xi $. Further, using Equation~\ref{eq:KLip}, $f(\delta_U) \leq \xi $ and positive $f(\delta_U)$ gives $|\delta^* - \delta_U| \leq (1/d) \xi$. 

A similar reasoning applies for $\delta^* < \delta_L$. We know that objective ${\sf BOPT}(\delta_L, S'', x''_S)$ with $\delta_L$ was positive (that is, $obj$ was negative) when $L$ was last updated. Note that we use $S'', x''_S \neq \overline{S}, \overline{x}_S$ as we assumed $U$ was the last update in binary search whereas $S'', x''_S$ is for when $L$ was last updated. As $\delta^* < \delta_L $, thus, due to decreasing $f$ we obtain $f(\delta^*) > f(\delta_L)$ meaning $f(\delta_L)$ is negative.  Using Equation~\ref{eq:xierror}, we get
$$
|f(\delta_L) - {\sf BOPT}(\delta_L, S'', x''_S) | \leq \xi
$$
With the knowledge that ${\sf BOPT}(\delta_L, S'', x''_S)$ is positive and $f({\delta_L})$ is negative, we can readily infer that $f(\delta_L) \geq -\xi $. Further, using Equation~\ref{eq:KLip}, $f(\delta_L) \geq -\xi $ and negative $f(\delta_L)$ gives $|\delta^* - \delta_L| \leq (1/d) \xi$.  

Recall the output of binary search is $\overline{\delta}_0, \overline{S}, \overline{x}_S$.
Combined with the possibility that $\delta^*$ can lie in $(\delta_L, \delta_U)$ we get the worst case error of choosing output $\overline{\delta}_0 = \delta_U$ (or $\overline{\delta}_0 = \delta_L$ in the analogous case) is $(1/d) \xi + \epsilon$.
Thus, we showed above that $|\delta^* - \overline{\delta}_0| \leq (1/d)\xi + \epsilon$. 

Also, $\overline{S}, \overline{x}_S$ is the approximate solution of problem $\sf BOPT$ with $\overline{\delta}_0$. Thus, $|f(\overline{\delta}_0) - {\sf BOPT}(\overline{\delta}_0, \overline{S}, \overline{x}_S)| \leq \xi$.
Further, $|f(\overline{\delta}_0) - f(\delta^*)| = |f(\overline{\delta}_0)| \leq D|\overline{\delta}_0 - \delta^*| \leq (D/d) \xi + D \epsilon$.
Combining these two facts we obtain:
$$
|{\sf BOPT}(\overline{\delta}_0, \overline{S}, \overline{x}_S)| \leq (D/d + 1) \xi + D \epsilon
$$
Note that ${\sf BOPT}(\overline{\delta}_0, \overline{S}, \overline{x}_S) = \sum_{j \in \overline{S}} N(\overline{x}_j) (w_j^d \overline{x}_j  + l_j^d ) - \overline{\delta}_0 D(\overline{x}_S)$. As $D(\overline{x}_S) > 0$, dividing throughout we get
$$
|\sum_{j \in \overline{S}} N(\overline{x}_j) (w_j^d \overline{x}_j  + l_j^d )/D(\overline{x}_S) - \overline{\delta}_0 | \leq [(D/d + 1) \xi + D \epsilon]/D(\overline{x}_S) 
$$
or using the notation $\cF$
$$
|\cF(\overline{S}, \overline{x}_S) - \overline{\delta}_0 | \leq [(D/d + 1) \xi + D \epsilon]/D(\overline{x}_S) 
$$
Combining, using triangle inequality, with the proved result that $|\delta^* - \overline{\delta}_0| \leq (1/d)\xi + \epsilon$ and the fact that $\delta^* = \cF(S^*, x^*_S)$ we get 
$$
|\cF(\overline{S}, \overline{x}_S) - \cF(S^*, x^*_S) | \leq (1/d)\xi + \epsilon + [(D/d + 1) \xi + D \epsilon]/D(\overline{x}_S) 
$$
Also, $D(x_S)$ is smallest when all $x$'s are 1 (as a worst case bound) giving $D(x_S) > \sum_{j \in S} e^{l^a_j}$. With a worst case over $S$, we have $D(x_S) > e^{\min_{j \in \cK}l^a_j}$ for any $S$. With the lower bound constant $L$ assumed, we get $D(x_S) > e^{L}$. Thus,
$$
|\cF(\overline{S}, \overline{x}_S) - \cF(S^*, x^*_S) | \leq [(1/d) + (D/d + 1)e^{-L}]\xi + [1 + De^{-L}]\epsilon
$$
The above is $O(\xi + \epsilon)$ as $d,D,L$ are constants. 

\textit{Details on Lipschitzness (intuitive and can be skipped by a time constrained reader but put here for completeness)}. Let $S', x'_S$ be optimal solution for $\delta'$. We can easily get that for any $\delta' > \delta$ (WLOG) we have 
$f(\delta)  - f(\delta') \geq \sum_{j \in S'} N(x'_j) (w_j^d x'_j  + l_j^d ) - \delta D(x'_S) - f(\delta') = D(x'_S)(\delta' - \delta)$, which using the fact proved above that $D(x'_S) > e^L$, we get $|f(\delta)  - f(\delta')| \geq e^L |\delta' - \delta|$, thus, $d = e^L$ (as $\delta, \delta'$ are arbitrary choices). 

Similarly, if $S, x_S$ is optimal solution for $\delta$ then $f(\delta)  - f(\delta') \leq f(\delta) - \sum_{j \in S} N(x_j) (w_j^d x_j  + l_j^d ) + \delta' D(x_S)  = (\delta' - \delta) D(x_S) $. Now, it is easy to show that $D(x_S)$ is largest when all $x$'s are 0 (as a worst case bound) giving $D(x_S) = \sum_{j \in S} e^{r^a_j}$. But, note that at max $C$ centers can operate, hence with a worst case over $S$, we have $D(x_S) < |C| e^{\max_{j \in \cK} r^a_j}$ for any $S$. But as $|C|$ and $\max_{j \in \cK} r^a_j$ are bounded, say by $C_0$ and $U_0$, this gives the Lipschitz constant $D =  C_0 e^{U_0}$
\end{proof}

\subsection{Proof of Theorem~\ref{thm:milpapxopt}}
\begin{proof} To analyze \eqref{prob:MILP-2}, we request the reader to first read the Appendix~\ref{sec:ilp} to get familiar with the problem \eqref{prob:MILP-2} introduced there. Let us denote $f^1(\theta,r,z)$ and $\cS^1$ be the objective function and feasible set of \eqref{prob:MILP-1}, and $f^2(\theta,r,z)$ and $\cS^2$ be the objective and feasible set of \eqref{prob:MILP-2}. We see that  $\cS^2 \subset \cS^1$ and $f^1(\theta,r,z) = f^2(\theta,r,z)$ for all $(\theta,r,z) \in \cS^2$, thus $\max_{(\theta,r,z) \in \cS^1}f^1(\theta,r,z) \geq \max_{(\theta,r,z) \in \cS^2}f^2(\theta,r,z)$. Now let $(\theta^*,r^*,z^*)$ be an optimal solution to \eqref{prob:MILP-1}, we create a solution $(\theta',r',z')$ such that $\theta' = \theta^*$, $z'_{jk} = r'_{jk} = 0$, $\forall k\in [K]$, if $\theta^*_j = 0$, $\forall j\in \cK$. We can see that $(\theta',r',z')$ is feasible to the both problems and $f^1(\theta^*,r^*,z^*) = f^1(\theta',r',z') = f^2(\theta',r',z')$. From the fact that $f^1(\theta^*,r^*,z^*)\geq \max_{(\theta,r,z) \in \cS^2}f^2(\theta,r,z) \geq f^2(\theta',r',z')$, we have that any optimal solution to \eqref{prob:MILP-2} is also optimal to \eqref{prob:MILP-1}, as desired. Since \eqref{prob:MILP-1} approximates \ref{prob:fraction-conversion}, then so does \eqref{prob:MILP-2}.

For the performance bound, let $H = \sum_{j} w^d_je^{\lambda r^a_j}(1+\lambda \max\{|l^d_j|;|l^d_j+w^d_j|\}) + \delta_0\lambda w^a_j e^{\lambda r^a}$.
Let $\widehat{N}(x_j)$ and $\widehat{g}_j$ be the piece wise linear approximations of $N(x_j)$ and $g_j(x_j)$. For any $x_j\in [0,1]$, let  $k\in [K]$ such that $x_j \in  [(k-1)/K, k/K]$, it can be seen that there is $\alpha \in [0,1]$ such that
\[
\widehat{N}(x_j)  = \alpha N((k-1)/K) + (1-\alpha) N((k-1)/K),
\]
which further implies that
\begin{align}
|\widehat{N}(x_j) - N(x_j)| &\leq \alpha |N(x_j) - N((k-1)/K)| \nonumber \\
&+ (1-\alpha)|N(x_j) - N(k/K)| \nonumber 
\end{align}
We then can use the mean value theorem to have the following inequalities
\begin{align}
|N(x_j) - &N((k-1)/K)| \leq 1/K\max_{x_j} |N'(x_j)| \nonumber \\
&=  1/K\max_{x_j} \left\{ \lambda w^a_j e^{\lambda(-w^a_j x_j + r^a_j)} \right\} \nonumber \\
&=  1/K \lambda w^a_j e^{\lambda r^a_j}  \nonumber 
\end{align}
We can have the same evaluation for $|N(x_j) - N(k/K)|$, which leads to 
\begin{equation}
\label{eq:approx-eq1}
    |\widehat{N}(x_j) - N(x_j)| \leq  1/K \lambda w^a_j e^{\lambda r^a_j}.  
\end{equation}
Similarly, we also have 
\begin{align}
    |g_j(x_j) - & g_j((k-1)/K)| \leq 1/K\max_{x_j} |g'(x_j)| \leq \psi \nonumber
\end{align}
where $\psi = \frac{1}{K}\left(w^d_j e^{\lambda r^a_j} + \lambda w^d_j e^{\lambda r^a_j } \max \{|l^d_j|; |l^d_j +w^d_j|\}\right)$. This also leads to 
\begin{equation}
\label{eq:approx-eq2}
    |\widehat{g}_j(x_j) - g_j(x_j)|\leq \psi.
\end{equation} 
Now, let $\widehat{F}(S,x_S)$ be the approximated objective function of ${B}(S,x_S)$, i.e., $\widehat{F}(S,x_S) = \widehat{g}_j(x_j) - \delta_0 \sum_{j} \widehat{N}(x_j) $. From the inequalities in \eqref{eq:approx-eq1} and \eqref{eq:approx-eq2} we have
\[
|\widehat{F}(S,x_S) -  {B}(S,x_S)|\leq H/K.
\]
Now, let $(S',x'_S)$ be an optimal solution to the approximate problem and $(S^{*},x^{*}_S)$ be an optimal solution to \eqref{prob:fraction-conversion}. We can write
{\small \begin{align}
|B(S',x'_S) - &B(S^*,x^*_S) | \leq  |B(S',x'_S) - \widehat{F}(S',x'_S)|  \nonumber \\
& +|\widehat{F}(S',x'_S) - B(S^*,x^*_S)| \nonumber \\
&\leq H/K +|B(S^{*},x^{*}_S) - \widehat{F}(S',x'_S)| \label{eq:approx-eq3}
\end{align}}
We further evaluate the second term of  \eqref{eq:approx-eq3} by  considering the  following two cases: $B(S^{*},x^{*}_S) \geq  \widehat{F}(S',x'_S)$ or $B(S^{*},x^{*}_S)\leq  \widehat{F}(S',x'_S) $. If $B(S^{*},x^{*}_S) \leq  \widehat{F}(S',x'_S)$ we have
\begin{align}
    |B(S^{*},x^{*}_S) - \widehat{F}(S',x'_S)| &=  \widehat{F}(S',x'_S) - B(S^{*},x^{*}_S)\nonumber \\
    &\leq \widehat{F}(S',x'_S) - B(S',x'_S) \nonumber \\
    &\leq H/K.
\end{align}
This can be done similarly for the second case to have  $|B(S^{*},x^{*}_S) - \widehat{F}(S',x'_S)|\leq H/K$. Combine this with \eqref{eq:approx-eq3} we obtain $|B(S^*,x^*_S) - B(S',x'_S) | \leq 2H/K$ as desired.

\end{proof}

\subsection{Proof of Lemma~\ref{closedformlemma}}
\begin{proof}
First, writing the solution in expanded form: 
$$
y^*_j = \Bigg\{ \begin{array}{lr l}
        e^{-\lambda w^a_j}, & & \text{for } 1 \leq  \beta_j\\
        e^{-\lambda w^a_j \beta_j}, & & \text{for } 0 < \beta_j < 1\\
        1, & & \text{for } \beta_j \leq 0
        \end{array}
$$
Next, the objective is
$$
 y_j e^{\lambda r^a_j} \big(\frac{-w_j^d \log y_j}{\lambda w_j^a}  + l_j^d - \delta_0 \big) + (\nu  + \mu_l)\frac{\log y_j}{\lambda w_j^a}
$$
While this objective is concave, we show that using the original formulation in $x$'s provides an easier path to a closed form solution. Transforming the variable back to $x$'s using $y_j = e^{-\lambda w^a_j x_j}$ for $0 \leq x_j \leq 1$ we get the objective as
$$
 e^{-\lambda w^a_j x_j + \lambda r^a_j} \big(w_j^d x_j  + l_j^d - \delta_0 \big) - (\nu  + \mu_l)x_j
$$
With a little manipulation, the above can be written as $ae^{-bx_j}x_j - cx_j + d$ for constants $a > 0,b > 0,c > 0, d$.
This function can be readily checked to be unimodal (one maximum and monotonic drop on either side) though not concave.
The first order condition for this is
$$
 e^{-\lambda w^a_j x_j + \lambda r^a_j} \big(w^d_j -\lambda w^a_j (w_j^d x_j  + l_j^d - \delta_0) \big) - (\nu  + \mu_l) = 0
$$
which is same as
$$
 w^d_j e^{-\lambda w^a_j x_j + \lambda r^a_j} \big( -\lambda w^a_j x_j +1 - \frac{\lambda w^a_j}{w^d_j}(l_j^d - \delta_0) \big) - (\nu  + \mu_l) = 0
$$
Using shorthand $z = -\lambda w^a_j x_j +1 - \frac{\lambda w^a_j}{w^d_j}(l_j^d - \delta_0)$, then
$$
 w^d_j e^{\lambda r^a_j + \frac{\lambda w^a_j}{w^d_j}(l_j^d - \delta_0) - 1}e^{z} \big( z \big) - (\nu  + \mu_l) = 0
$$
which gives, by definition of Lambert W function~\cite{corless1996lambertw},
$$
z = W\Big(\frac{\nu  + \mu_l}{w^d_j} e^{1-\lambda r^a_j - \frac{\lambda w^a_j}{w^d_j}(l_j^d - \delta_0)}\Big)
$$
Then, using the definition of $z$, we solve for $x_j$. Denoting
$$
\beta = \frac{1}{\lambda w^a_j }\Big[ 1- \frac{\lambda w^a_j}{w^d_j}(l_j^d - \delta_0) - W\big(\frac{\nu  + \mu_l}{w^d_j} e^{1-\lambda r^a_j - \frac{\lambda w^a_j}{w^d_j}(l_j^d - \delta_0)}\big) \Big]
$$
the solution gives $x_j = \beta$. However, $x_j$ is constrained to lie in $[0,1]$ and using the fact that the objective is unimodal, we can state that 
if $\beta$ is more than $1$ then $x_j = 1$ and if $\beta$ is less than $0$ then $x_j = 0$. This gives the required result.
\end{proof}

\subsection{Proof of Lemma~\ref{fixeddualssolverproperty}}
\begin{proof}For part(1), all expressions in the algorithm can be evaluated exactly, except for the solution of the optimization in line 4. But, the optimization in line 4 can be computed by the closed form formula given in Lemma~\ref{closedformlemma}. The closed form requires computing the Lambert W function, which is possible with arbitrary constant precision in $O(1)$ evaluations of the exponential function~\cite{johansson2020computing}. This gives us the desired result.

For part(2), the optimization in line 4 can be computed by $O(1)$ evaluations of the exponential function~\cite{johansson2020computing}, which is overall $O(1)$ as exponentiation is assumed constant time. All loops run for a maximum of $|\cK|$ iterations, and the operations in them are all constant time. Other than the loops the step are sorting $H$ (of length $|\cK|$) and  partitioning it. The runtime is dominated by the sorting of $|\cK|$ sized array, which gives $O(|\cK| \log |\cK|)$.
\end{proof}


\subsection{Proof of Theorem~\ref{th:hybrid}}
\begin{proof}
Recall $\Phi(S,\nu, \mu,\delta_0) = \max_{y_S \in \cD} \phi(S,\nu, \mu, y_S, \delta_0)$ (the objective function of ${\sf SwitchedDualOPT}$). For ${\sf BOPT}$, we want to solve ${\sf DualOPT}$ but instead we solve ${\sf SwitchedDualOPT}$ and that too approximately with additive error $\xi$. However, the first half of the proof of Theorem~\ref{binsearchthm} does not depend on what ${\sf BOPT}$ is except for the result that $f$ (which is ${\sf SwitchedDualOPT}$ here when solved exactly) is monotonic in $\delta$. 

In order to run the heuristic ${\sf SwitchedDualOPT}$ within Algorithm \ref{overallsolver}, it is necessary to have the objective value of ${\sf SwitchedDualOPT}$ be monotonically decreasing in $\delta_0$. To prove this, we first note that if $u(y),u'(y)$ are two functions from a compact set $Y$ to $\mathbb{R}$, then if $u(y) \geq u'(y)$ for all $y\in Y$, then we have
\begin{align}
    \max_{y\in Y} u(y) \geq \max_{y\in Y} u'(y) \nonumber\\
    \min_{y\in Y} u(y) \geq \min_{y\in Y} u'(y). \nonumber
\end{align}
To verify the above, let $y^*$ be an optimal solution to $\max_{y\in Y} u'(y)$. We then have $\max_{y\in Y} u'(y) = u'(y^*) \leq u(y^*) \leq \max_y u(y)$. The second inequality can be verified similarly by letting $y^{**} = \text{argmin}_y  u(y) $ and have  $\min_{y\in Y} u(y) = u(y^{**}) \geq u'(y^{**}) \geq \min_{y} u'(y)$.

We now come back to the objective function of ${\sf SwitchedDualOPT}$. Given $\delta_1 \geq \delta_2$, for any $S,\nu,\mu,y_S$ we have 
$\phi(S,\nu, \mu, y_S,\delta_1)\leq \phi(S,\nu, \mu, y_S,\delta_2)$. The above remark tells us that
{\small\begin{align}
    \max_{y_S} \phi(S,\nu, \mu, y_S,\delta_1)&\leq \max_{y_S}\phi(S,\nu, \mu, y_S,\delta_2) \nonumber \\
        \max_S \max_{y_S} \phi(S,\nu, \mu, y_S,\delta_1)&\leq \max_S \max_{y_S}\phi(S,\nu, \mu, y_S,\delta_2) \nonumber \\
        \min_{\nu,\mu}\max_S \max_{y_S} \phi(S,\nu, \mu, y_S,\delta_1)&\nonumber \\
        \leq \min_{\nu,\mu}\max_S \max_{y_S}&\phi(S,\nu, \mu, y_S,\delta_2), \nonumber
\end{align}}
which implies the monotonicity of the objective function of ${\sf SwitchedDualOPT}$, as desired.

The Lipschitz constants can be derived in same manner as Theorem~\ref{binsearchthm} and for notational simplicity we call them $d, D$ here again. In this part, we choose small $\xi$ such that $(1/d)\xi \leq \epsilon$. Thus, using the partial result from Theorem~\ref{binsearchthm} that $|\overline{\delta}_0 - \delta^*|\leq (1/d)\xi + \epsilon$ we get $|\overline{\delta}_0 - \delta^*| \leq 2\epsilon$.


Recall that $f(\delta^*) = 0$. Thus, after running Algorithm \ref{overallsolver} with $f = {\sf SwitchedDualOPT}$ we will obtain $\overline{\delta}_0$ such that 
\[
  \min_{\nu,\mu} \max_{S} \left\{\Phi({S},\nu, \mu,\overline{\delta}_0+2\epsilon)\right\}  \leq 0.
\]
Moreover, from the minimax inequality (which always holds) we have $$\min_{\nu,\mu} \max_{S} \left\{\Phi({S},\nu, \mu,\overline{\delta}_0+2\epsilon)\right\} 
\geq  \max_{S} \min_{\nu,\mu} \left\{\Phi({S},\nu, \mu,\overline{\delta}_0+2\epsilon)\right\},$$ thus 
\[
\max_{S} \min_{\nu,\mu} \left\{\Phi({S},\nu, \mu,\overline{\delta}_0+2\epsilon)\right\} \leq 0,
\]
The above is ${\sf DualOPT}$ which has same optimal as ${\sf BOPT}$. Then, if $S^*, x^*_S$ is the optimal solution for ${\sf EqOPT}$, the above leads to the inequality 
\[
\cF (S^*,x^*_S) \leq  \overline{\delta}_0+2\epsilon.
\]
On the other hand, clearly $\cF (S^*,x^*_S) \geq \cF(\overline{S},\overline{x}_S)$. 
This directly gives the upper and lower bound.

The inequality 
\[
\cF (S^*,x^*_S) \leq  \overline{\delta}_0+2\epsilon.
\]
also gives
\begin{align*}
  \cF (S^*,x^*_S) - \cF (\overline{S},\overline{x}_S) \leq  &\; \overline{\delta}_0 - \cF(\overline{S},\overline{x}_S) +2\epsilon \\
  \leq &\; |\overline{\delta}_0 - \cF(\overline{S},\overline{x}_S)| +2\epsilon.  
\end{align*}

\end{proof}

\end{document}